\documentclass[fleqn,usenatbib,useAMS]{mnras}

\usepackage{newtxtext,newtxmath}

\usepackage[T1]{fontenc}
\usepackage{ae,aecompl}

\usepackage{float}
\usepackage{graphicx}	
\usepackage{amsmath}	
\usepackage{multicol}        
\usepackage{bm}		

\usepackage{xcolor}
\usepackage{soul}

\newcommand{\pcc}{{\rm \,cm}^{-3}}

\title[Magnetic fields in Pop~III star formation]{Impact of magnetic fields on Population III star formation}

\author[C. R. Saad et al.]{
Cynthia R. Saad,$^{1}$\thanks{E-mail: crs07@mail.aub.edu}
Volker Bromm$^{2}$ and
Mounib El Eid$^{1}$
\\
$^{1}$Department of Physics, American University of Beirut, PO Box 11-0236, Riad El-Solh, Beirut 11097 2020, Lebanon\\
$^{2}$Department of Astronomy, University of Texas, Austin, TX 78712, USA\\
}

\date{Accepted XXX. Received YYY; in original form ZZZ}

\pubyear{2022}

\begin{document}
\label{firstpage}
\pagerange{\pageref{firstpage}--\pageref{lastpage}}
\maketitle

\begin{abstract}
The theory of the formation of the first stars in the Universe, the so-called Population~III (Pop~III), has until now largely neglected the impact of magnetic fields. Complementing a series of recent studies of the magneto-hydrodynamic (MHD) aspects of Pop~III star formation, we here carry out a suite of idealized numerical experiments where we ascertain how the fragmentation properties of primordial protostellar discs are modified if MHD effects are present. Specifically, starting from cosmological initial conditions, we focus on the central region in a select minihalo at redshift $z\sim 25$, inserting a magnetic field at an intermediate evolutionary stage, normalized to a fraction of the equipartition value. To explore parameter space, we consider different field geometries, including uniform, radial, toroidal, and poloidal field configurations, with the toroidal configuration being the most realistic. The collapse of the gas is followed for $\sim$8 orders of magnitude in density after the field was inserted, until a maximum of $10^{15} \pcc$ is reached. We find that the magnetic field leads to a delay in the collapse of the gas. Moreover, the toroidal field has the strongest effect on the collapse as it inhibits the fragmentation of the emerging disc surrounding the central core and leads to the formation of a more massive core. The full understanding of the formation of Pop~III stars and their mass distribution thus needs to take into account the effect of magnetic fields. We further conclude that ideal MHD is only a first step in this endeavor, to be followed-up with a comprehensive treatment of dissipative effects, such as ambipolar diffusion and Ohmic dissipation.
\end{abstract}




\section{Introduction}

The formation of the first stars in the Universe, the so-called Population~III (Pop~III), represents a crucial transition in cosmic history, where the simple initial conditions left behind by the inflationary fireball gave rise to an ever increasing complexity \citep[][]{barkana2001, bromm2004, bromm2009,loeb2010}. In particular, Pop~III stars cannot be composed of elements beyond lithium according to the well-established Big Bang nucleosynthesis (see \citealt{makki2019}, and references therein). Understanding their formation has been a challenging task in astrophysics and cosmology, and has in particular relied on numerical simulations \citep[reviewed in][]{bromm2013}. One key motivation has been to provide a framework for observations with the recently-launched James Webb Space Telescope (JWST), which will soon explore the earliest structures in the Universe.

Since there are no direct observations of these first stars yet, one can only rely on a theoretical approach in order to characterize their properties. High-precision observations of CMB temperature fluctuations have provided a detailed picture of the initial conditions for cosmological structure formation. First discovered by the Cosmic Background Explorer (COBE) mission (e.g. \citealt{smoot1992}), the CMB anisotropies were subsequently scrutinized by the Wilkinson Microwave Anisotropy Probe (WMAP) with greatly improved angular resolution (e.g. \citealt{wmap2011}). The {\it Planck} mission further reduced the uncertainties due to dust contamination, owing to its ten times better sensitivity and two times better angular resolution (e.g. \citealt{cmb2016}). These observations have shown that the Universe was initially almost homogeneous with very small density perturbation of order $10^{-5}$, equivalent to temperature fluctuations of 0.0002\,K from point to point across the sky.

The initial conditions under which the first stars form are given by the $\Lambda$CDM model of structure formation, with parameters provided by the CMB measurements. These conditions are different from the contemporary star formation process \citep[e.g.][]{McKee2007}, yet not as simple as it was once assumed. Actually, the chemistry network is indeed relatively simple due to the lack of heavy elements, and any radiative feedback from the surrounding, since these stars are the first to ever form in an otherwise dark Universe. However, several studies have elaborated the complexity involved by considering the radiation-hydrodynamical feedback from the accreting protostar (e.g. \citealt{wise2012}, \citealt{hirano2014}), dark matter annihilation (e.g. \citealt{stacy2014}), as well as the primordial streaming velocities (e.g. \citealt{stacy2011a,Schauer2021}). In addition, including the effect of magnetic fields, to be discussed later, renders the problem more complex and challenging.

The particle physics nature of dark matter (DM) is an essential ingredient in our understanding of first star formation, setting the cosmological environment for the dissipative collapse of the primordial gas into DM minihaloes, until reaching a number density of $10^4\pcc$, where the gas decouples from the DM and becomes self-gravitating \citep{bromm2002}. The evolution and final stages of Pop~III stars depend crucially on their achieved mass, stellar rotation, and stellar multiplicity (see reviews by \citealt{bromm2013,bromm2020}, and references therein), with the stellar mass being the most crucial parameter, since it determines the star’s luminosity, lifetime, end stage and chemical footprint. Although this is subject to vigorous debate, one can argue that for the first stars, final masses are close to initial masses, due to the lack of radiatively-driven mass loss. In addition, other mass-loss mechanisms may be suppressed or absent as well. E.g., as argued by \cite{baraffe2001}, very massive zero-metal stars of masses in excess of $M_{*}=120\,M_{\odot}$ are unstable to nuclear-powered radial pulsations on the main sequence, nevertheless, they do not suffer appreciable mass loss because the growth timescale of the pulsational instability is much longer than for metal-rich counterparts. 

A key open issue is the initial mass function (IMF) of Pop~III stars, determining their evolutionary pathways and fates. In particular, as mentioned above, if mass loss will not reduce the initial mass effectively, the IMF may extend to the $M_{*}=140-260\,M_{\odot}$ range for the occurrence of a peculiar supernova triggered by electron positron pair creation, the so-called pair-creation SN (PCSN), powered by explosive oxygen burning \citep[e.g.][]{ober1983, heger2002}. Outside this range, black hole formation would be the end stage. A successful PCSN leads to total disruption of the star and to chemical enrichment only up to the iron group elements. On the other end of the spectrum, if Pop~III stars would have been formed with sub-solar masses, they would have survived to the present-day (see \citealt{heger2003}). However, such Pop~III fossils have never been observed \citep[e.g.][]{hartwig2015}. Prior to $\sim$2010, the consensus view was that Pop~III stars formed as single massive stars (e.g. \citealt{bromm2002}, \citealt{yoshida2003}, \citealt{abel2002}, \citealt{oshea2007}). To the contrary, more recent works (e.g. \citealt{turk2009}, \citealt{stacy2010}, \citealt{clark2011b}, \citealt{greif2011b}, \citealt{stacy2013b}) argue that due to limited numerical resolution, the earlier investigations could not resolve the possible fragmentation within the protostellar accretion disc into lower-mass objects. Despite this sub-fragmentation, the IMF is still predicted to be top-heavy, with an uncertain upper mass limit \citep[e.g.][]{stacy2016}. Clearly, the mass range of Pop~III stars is still a matter of debate.

Early studies of Pop~III star formation did not include magnetic fields, assuming that any primordial seed field would not be dynamically important \citep[e.g][]{glover2013}. The origin of cosmic magnetic fields is not well known. E.g., \cite{jedamzik2020} suggested that a pre-recombination primordial magnetic field of strength $\sim 0.1 $\,nG could explain the origin of extragalactic magnetic fields. Several mechanisms have been presented for the generation of seed fields (reviews by \citealt{kandus2011}, \citealt{subramanian2016}). Examples are primordial vorticity during the radiation era of the early Universe (e.g. \citealt{harrison1970}), the quark-hadron phase transition (e.g. \citealt{bonometto1993}), or the electroweak phase transition (e.g. \citealt{baym1996}). A popular idea is the creation of cosmic magnetic fields during the inflationary epoch or later via the so-called ``Biermann battery'' (e.g. \citealt{biermann1950}, \citealt{xu2008}). Such seed fields could be quickly amplified to equipartition with the kinetic energy in protogalactic clouds due to turbulent small-scale dynamo action (e.g. \citealt{kulsrud1997}, \citealt{schober2012}, \citealt{mckee2020}). Possible observational probes target the strength and spatial distribution of magnetic fields in galaxies, the IGM, or the inter-cluster medium (ICM), and in objects at high redshifts \citep[reviewed in][]{grasso2001}. These observations are rotation measure (RM) determinations based on the Faraday rotation of polarized electromagnetic radiation passing through an ionized medium (see \citealt{kronberg1994}), finding evidence for maximum field strengths of $\sim 1 \mu$G in clouds at $z\gtrsim 1$.

Several authors have pioneered the investigation of magneto-hydrodynamical (MHD) effects on the formation of Pop~III stars (e.g. \citealt{machida2008}, \citealt{schleicher2009}, \citealt{schleicher2010}, \citealt{turk2012}, \citealt{schober2012}, \citealt{latif2016}, \citealt{hirano2017}), followed recently by more physically realistic studies (e.g. \citealt{mckee2020}, \citealt{garaldi2021}).
In the work by \cite{sharda2020a}, isolated, initially-turbulent primordial cores with different initial field strengths were considered, finding that the magnetic field inhibits fragmentation and favors the formation of single stars of high masses. In a subsequent paper, \cite{sharda2021} found a strong dependence on the initial conditions. Similarly, \cite{sadanari2021} studied the collapse of magnetized primordial gas cores with self-consistent thermal evolution and found that the magnetic fields slow down the cloud contraction only in the directions perpendicular to the field lines but not affecting the temperature evolution of the central core. In addition, these authors argued that magnetic braking leads to efficient transport of angular momentum which inhibits fragmentation, above a critical magnetic field strength of $B > 10^{-8} (n_H/1 \pcc)^{2/3}$\,G during the collapse phase. 

Starting from cosmological initial conditions, \cite{koh2021} noticed that the magnetic field delayed the gravitational collapse by $\Delta z = 2.5$, or completely suppressed collapse in minihaloes for sufficiently high field strengths. Similarly, \cite{stacy2022} compared the results of cosmological simulations with and without magnetic fields included. Introducing a field with an initial strength just above the value predicted from the Biermann battery, they found that this seed field was amplified to about half the equipartition value at number density $n=10^8\pcc$. The main effect of the magnetic field was to suppress fragmentation. In the pure hydrodynamic (HD) case, several protostars formed with masses ranging from 1 to 30\,$M_{\odot}$, whereas only a single $\sim$ 30\,$M_{\odot}$ protostar formed by the end of the MHD simulation. In contrast, recent work by \citet{prole2022} argues that the magnetic field is unimportant for densities below $10^8\pcc$. In fact, this study introduces a saturated magnetic field when the central density reaches $\sim 10^{10}\pcc$, without self-consistently resolving its prior amplification. The authors conclude that the number of sink particles formed, and the total mass accreted, were not affected by the magnetic field. Hence, fragmentation and the resulting IMF would not be impacted by MHD effects, as was argued in previous studies, thus demonstrating the vigorous state of this ongoing debate.

Here, we investigate the effect of magnetic fields on the formation of Pop~III stars with a complementary numerical strategy. We also start from cosmological initial conditions, but introduce magnetic fields in an idealized fashion. In difference from previous studies, we do not attempt to follow the amplification of the seed field through small-scale turbulent dynamo action, but instead introduce the fully developed field at a later stage, in accordance with analytical models of the expected equipartition process (see below). This idealized approach is motivated by limitations on computational resources that cannot resolve the true character of turbulence, particularly the vortical motions at small scales that are ultimately responsible for magnetic field amplification \citep[][]{sur2010,federrath2011,turk2012}.

In Section~\ref{sect2}, we present our numerical methodology, including the equations that govern the dynamics, the chemical and thermal evolution of the gas, the adopted initial conditions, as well as the implementation of the magnetic field, its magnitude and its configurations. Section~\ref{sect3} contains the discussion of our results, and we summarize and offer conclusions in Section~\ref{sect4}.

\section{Numerical Methodology}
\label{sect2}

\subsection{Basic equations}
\label{sect2.1}
We carry out a suite of cosmological simulations, with the standard physics of DM and gas (baryons) included, employing the three-dimensional numerical code \texttt{Enzo}, released by \cite{enzo}. This code has the advantage of combining an N-body solver with adaptive mesh refinement (AMR) hydrodynamics, within the context of cosmological initial conditions. In order to resolve fragmentation, a spatial resolution of 64 grid cells per Jeans length is adopted, satisfying the Truelove criterion for avoiding artificial fragmentation \citep{truelove1997}.
\texttt{Enzo} implements the standard equations of MHD, expressing conservation of mass, momentum, and energy, as well as the coupling between magnetic fields and matter (e.g. \citealt{thorne2017}).\\
Since the simulations are started from cosmological initial conditions, the standard MHD equations need to be modified for an expanding Universe (e.g. \citealt{enzo}).
In this formulation, the comoving quantities that are evolved by the solver are related to the proper observable quantities by the following equations:
\begin{gather}
    \rho_{\mathrm{proper}}=\rho a^{-3}\mbox{\ ,}\\
    p_{\mathrm{proper}}=p a^{-3}\mbox{\ ,}\\
    \bm{v}_{\mathrm{proper}}=\bm{v}+\dot{a}\bm{x}\mbox{\ ,}\\
    \Phi_{\mathrm{proper}}=\Phi-\frac{1}{2}a\Ddot{a}\bm{x}^2\mbox{\ ,}\\
    E=a^3 \left( E_{\mathrm{proper}}-\dot{a}\bm{x}\cdot \bm{v}_{\mathrm{proper}}-\frac{1}{2}\dot{a}^2 \bm{x}^2\right) \mbox{\ ,}\\
    \bm{B}_{\mathrm{proper}}=\bm{B}a^{-3/2}\mbox{\ ,}\label{Eq.Bprop}
\end{gather}
where $\rho$ is the comoving density, $a = 1/(1 + z)$ the expansion factor, $p$ the gas pressure, and $\bm{v}$ the velocity in the comoving frame. Further, $\bm{x}$ gives the position of the baryons in the comoving frame, $\Phi$ is the comoving gravitational potential from both dark matter and baryons, $E$ the comoving total fluid energy density, and $\bm{B}$ the comoving magnetic field. The scaling in Equ.~\ref{Eq.Bprop} is applicable when using the Dedner divergence cleaning method for enforcing the solenoidal constraint \citep[see section~4.2 in][]{enzo}.\\
The comoving total fluid energy density and total pressure are given by:
\begin{gather}
    E = \frac{\rho{v}^2}{2}+\frac{p}{\gamma - 1}+\frac{B^2}{8\pi}\mbox{\ ,} \\
    \Tilde{p}=p+\frac{B^2}{8\pi}\mbox{\ .}
\end{gather}
Here, $\gamma$ is the adiabatic index (the ratio of the specific heats), and $\Tilde{p}$ the total comoving pressure, representing the sum of the thermal and magnetic pressures.

The adiabatic index, which depends on the chemical composition and gas temperature, is given by \cite{omukai1998}:
\begin{equation}
    \gamma_{\rm ad}=1+\Sigma n(i)/\Sigma \frac{n(i)}{\gamma_i -1}\mbox{\ .}
\end{equation}
For monatomic species, the adiabetic index is $\gamma_i = 5/3$, whereas for H$_2$, rotational and vibrational degrees of freedom have to be taken into account as follows:
\begin{equation}
    \frac{1}{\gamma_{\rm H_2}-1}=\frac{1}{2}\left[5+2x^2\frac{e^x}{(e^x-1)^2}\right]\mbox{\ ,}
    \label{gammah2}
\end{equation}
where $x=6100{\rm \,K}/T$.

\subsection{Primordial chemical network}
\label{sect2.2}
In the present calculations, a modified version of the \texttt{Grackle}\footnote{\url{https://grackle.readthedocs.io/}} chemistry and cooling library (e.g. \citealt{grackle}) is employed to solve the non-equilibrium chemistry network of the 12 primordial species (H, H$^+$, H$^-$, H$_2$, H$_2^+$, He, He$^+$, He$^{++}$, D, D$^+$, HD, $e^-$). These species are linked by 33 reactions including the formation of molecular hydrogen via the H$^-$ and H$_2^+$ channels \citep[][]{tegmark1997}, and also via three-body reactions. In addition, the key cooling and heating processes are incorporated, such as H$_2$ ro-vibrational transitions, chemical heating and induced emission. It is well-known that the H$_2$ formation pathway is sensitive to the three-body reaction rate (e.g. \citealt{turk2010}), with a concomitant impact on the Pop~III star formation process. The specific rate employed in these simulations is the one given in \cite{palla1983}.

\texttt{Grackle} evolves the Lagrangian energy equation and solves the stiff network of coupled chemical rate equations with a low-order backwards difference formula (BDF) approach, due to its stability and ease of implementation \citep{anninos1997}. With the total cooling and heating rates in hand, we can write:
\begin{equation}
    \frac{\rm d \textit{e}}{\rm dt}=-\dot{e}_{\rm cool}+\dot{e}_{\rm heat}\mbox{\ .}
\end{equation}
Here, $e$ is the specific internal energy, related to temperature via $e=k_\text{B}T/[(\gamma-1)\mu m_\text{H}]$, with $k_\text{B}$ being the Boltzmann constant, $\mu$ the mean molecular weight, and $m_\text{H}$ the mass of a hydrogen atom. 

The integrator is sub-cycled according to a time-step constraint, to enhance accuracy:
\begin{equation}
    \Delta t \leq 0.1 \frac{e}{\dot{e}}\mbox{\ .}
\end{equation}
To solve the rate of change for a given chemical species, the creation and destruction rates are grouped as follows:
\begin{equation}
    \frac{\partial n_{\rm i}}{\partial t}=C_{\rm i}(T,n_{\rm j})-D_{\rm i}(T,n_{\rm j})n_{\rm i}\mbox{\ ,}
    \label{cd}
\end{equation}
where $C_{\rm i}$ is the total creation rate of the i-th species for a given temperature $T$ and the abundances of the other species. The second term on the right-hand side of equation~(\ref{cd}) is the destruction rate of the given species.

Following the collapse to the stage of a central hydrostatic (protostellar) core would require reaching number densities of $n\gtrsim 10^{20} \pcc$, where the primordial gas becomes opaque to its cooling radiation \citep{Rees1976}. Although reaching such stellar densities has been achieved with cosmological simulations of extreme dynamical range \citep{greif2012}, the challenge then is to follow the accretion process onto the growing core for $\sim 10^4$\,yr, the Kelvin-Helmholtz timescale for massive stars \citep[e.g.][]{Omukai2003}. With the requirement on the time step imposed by the Courant condition, such a numerical {\it tour de force} is still not feasible \citep{greif2015}. A widely used numerical compromise is to insert sink particles at somewhat lower densities, where the `Courant myopia' stops being computationally prohibitive \citep[see][for an implementation within the AMR methodology]{Safranek2016}. Another powerful idealization is the escape probability method combined with the Sobolev approximation to treat the complex radiative transfer of partially opaque H$_2$ line cooling (e.g. \citealt{yoshida2008}, \citealt{clark2011b}, \citealt{greif2011b}).

We here adopt the alternative approach of \cite{hirano2017}, who introduced an artificially stiffened equation of state at a threshold density, $n_\text{th}$, chosen for reasons of computational expediency. This stiffening is implemented through an artificially imposed optical depth at $\sim n_\text{th}$, resulting in an artificial hydrostatic core inside the collapsing minihalo. This artificial core is larger and more massive than the true protostellar core, representing the resolution limit of the simulation. In this work, the radiative cooling for gas elements with density exceeding a threshold number densities of $n_{\rm th} = 10^{12}\pcc$ is reduced. At number densities $n_\text{H} \goa 10^{12}\pcc$, the primordial gas begins to become optically thick to H$_2$ line emission, but other cooling channels, such as collision-induced emission, will remain effective until much higher densities \citep{greif2015}, beyond our artificial threshold density. We therefore do not accurately resolve the star formation process on the smallest scales. Specifically, we impose an artificial optical depth via \citep{hirano2017}:
\begin{equation}
\label{Eq.tauart}
    \tau_{\rm art} = \left(\frac{n_{\rm H}}{n_{\rm th}}\right)^2 \mbox{\ .}
\end{equation} 
The corresponding escape fraction is: 
\begin{equation}
\label{Eq.Betaart}
\beta_{\rm esc,art} = \frac{1-\exp{(- \tau_{\rm art})}}{\tau_{\rm art}}\mbox{\ .}
\end{equation}
All radiative cooling rates are reduced by this factor. This treatment has the advantage that the complicated hydrodynamics inside the opaque core does not need to be followed. Similarly, solving the energy and chemical rate equations at the increasingly high densities beyond the resolution scale is now avoided, replaced by a simple adiabatic evolution.

To test for numerical convergence, we consider select cases with an increased threshold number density of $n_{\rm th} = 10^{15}\pcc$. We evolve these high-resolution runs as far as computationally possible, given the now much shorter required Courant time steps. To approximately account for the increased optical depth of the primordial gas to H$_2$ line cooling in the high-resolution runs at $n \gtrsim 10^{12}\pcc$ \citep{omukai1998, ripamonti2002}, the H$_2$ cooling rate is decreased by a density-dependent term in \texttt{Grackle}, as expressed in equation~19 in \citet{ripamonti2004}.

Determining the extent of the protostar in simulations that do not insert sink particles can be achieved by finding the photospheric surface of the protostar, corresponding to the location where the optical depth reaches unity \citep[e.g., fig.~9 in][]{Stacy_structure2013}. In our idealized model, the effective size of a `protostar' is given by the surface where the artificial opacity is unity, i.e. where the number density reaches the threshold density of $n_{\rm th} = 10^{12}\pcc$. Again, we emphasize that we here cannot resolve the true protostar, in terms of mass and radius, and this should be kept in mind when considering the resulting fragment masses and mass distribution (see below).

\begin{figure}
\centering
\begin{multicols}{2}
    \includegraphics[width=0.48\textwidth]{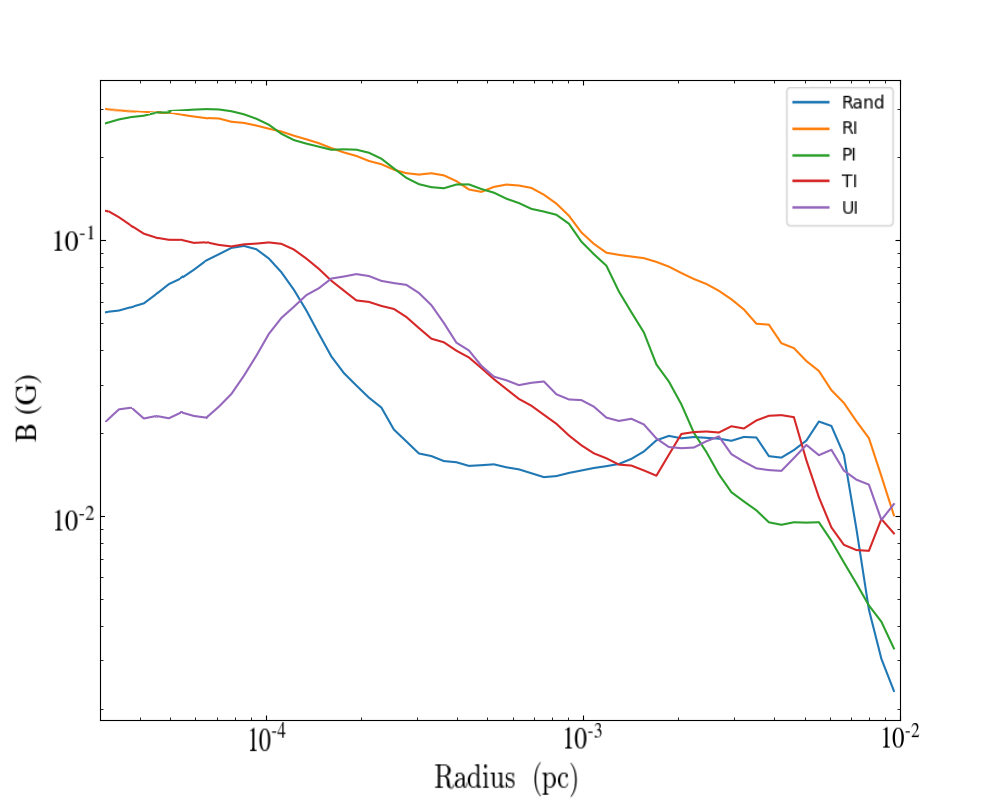}
    \includegraphics[width=0.48\textwidth]{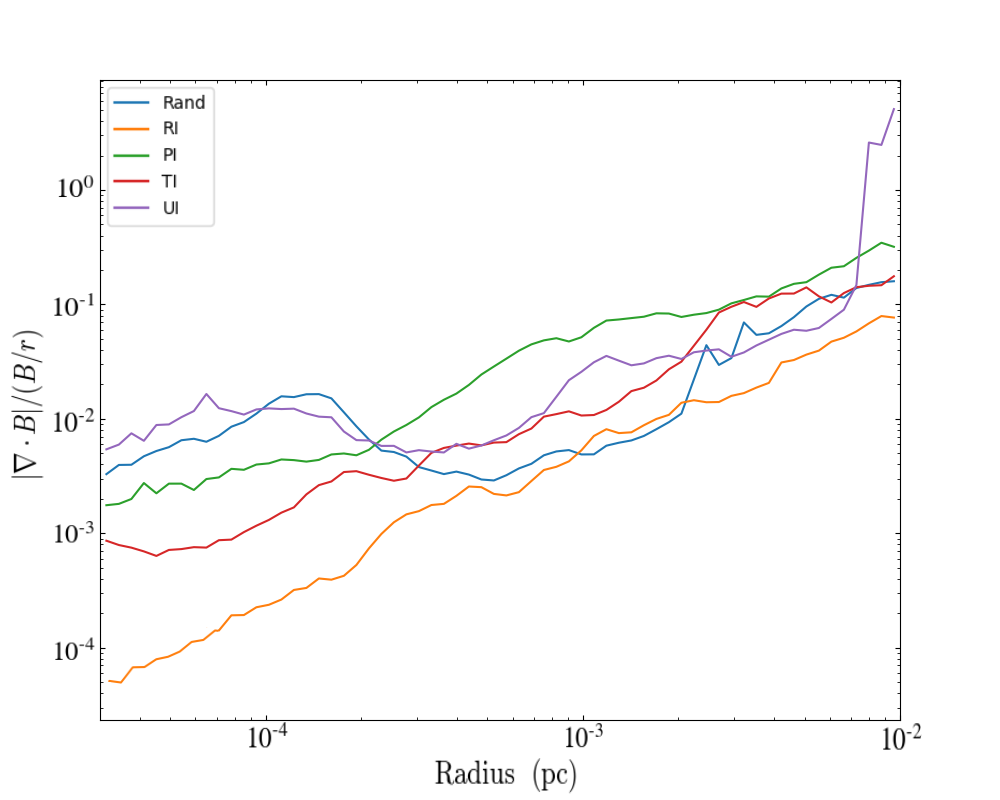}
\end{multicols}
\caption{Magnetic field properties in the center of the star formation region, evaluated when the gas density reaches a maximum value of $10^{12} \pcc$. {\it Top panel:} Magnitude of the magnetic field, according to equation~(\ref{Eq.Bsat}) with $\eta = 0.5$, as a function of radius for different field geometries, as indicated. {\it Bottom panel:} Magnitude of the magnetic field divergence, expressed in dimensionless form, vs. radius. As can be seen, the $\bm{\nabla} \cdot \bm{B} =0$ constraint is well satisfied throughout, which is required for the MHD calculation.}
\label{fig:comparisonB}
\end{figure}
\begin{figure*}
\begin{multicols}{3}
    \includegraphics[width=0.3\textwidth]{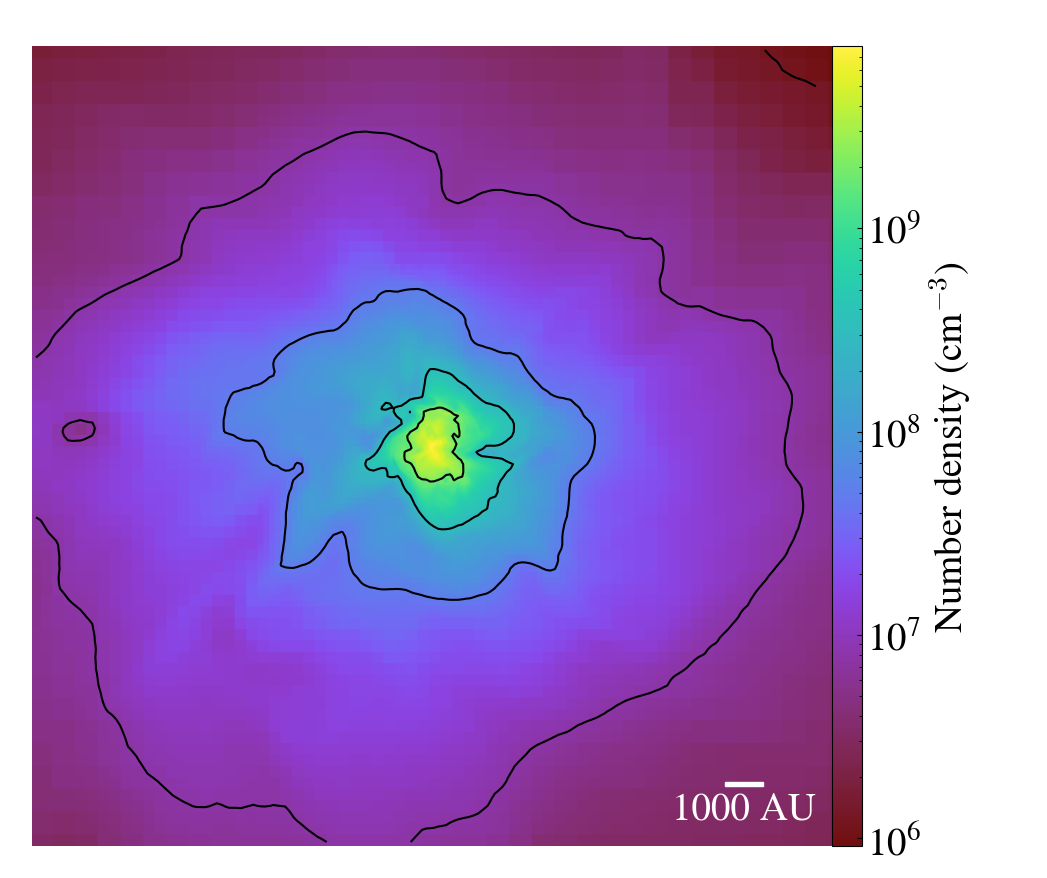}\par
    \includegraphics[width=0.3\textwidth]{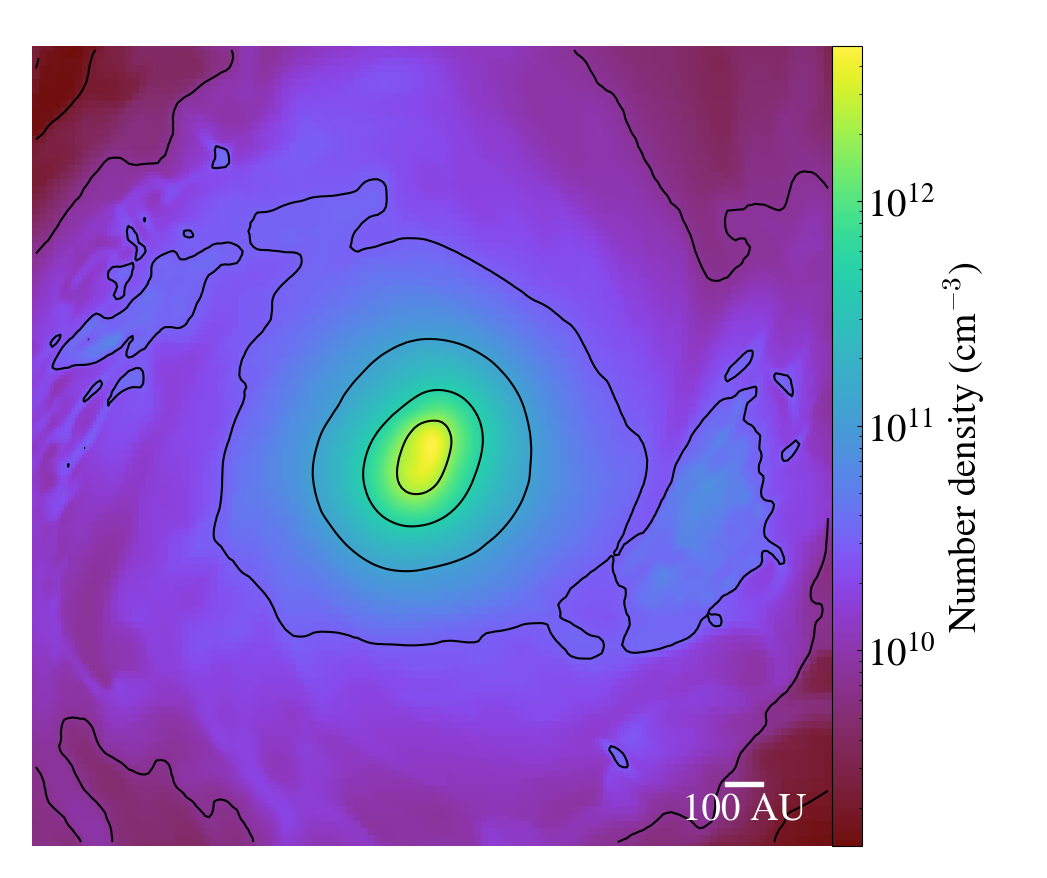}\par
    \includegraphics[width=0.3\textwidth]{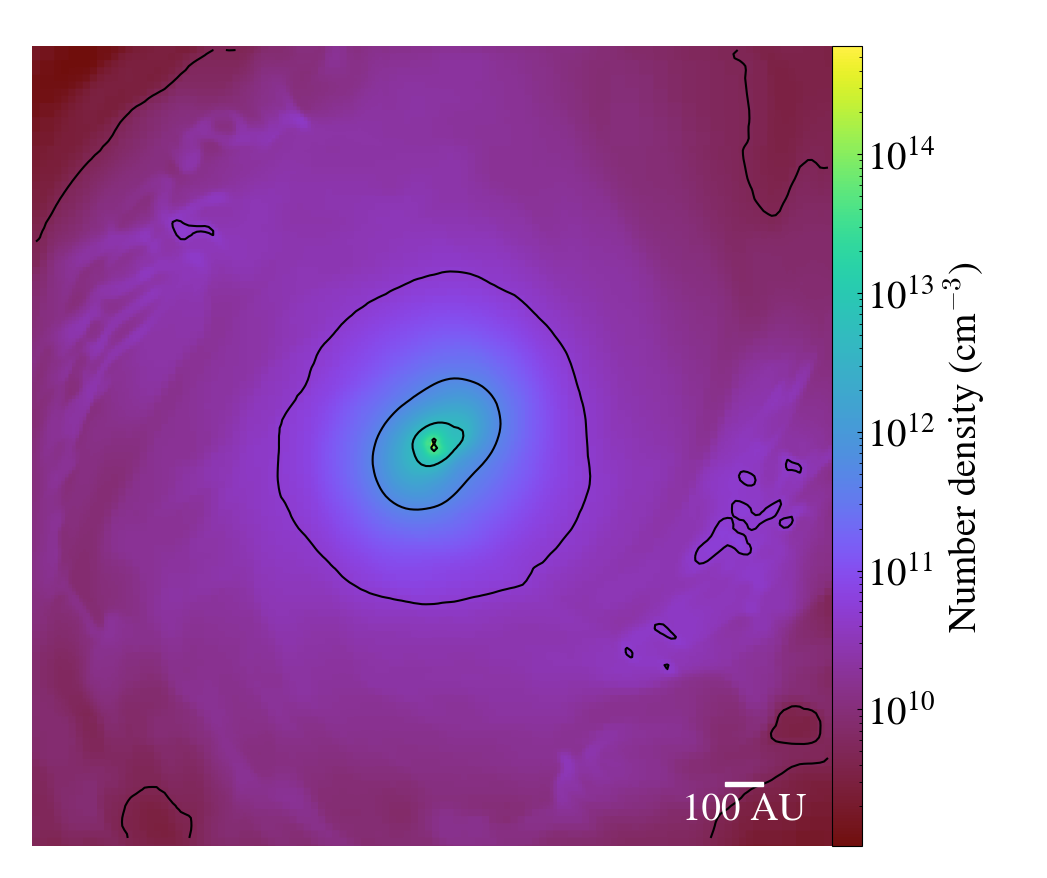}
\end{multicols}
\begin{multicols}{3}
    \includegraphics[width=0.3\textwidth]{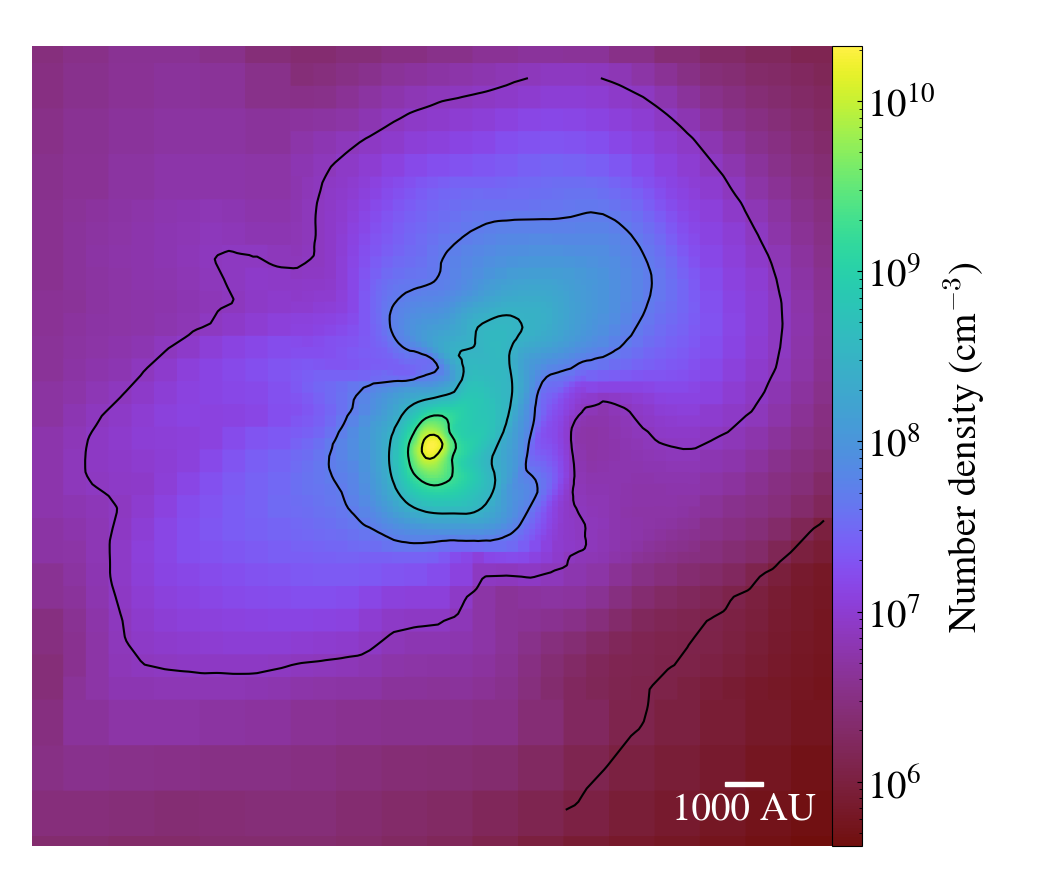}\par
    \includegraphics[width=0.3\textwidth]{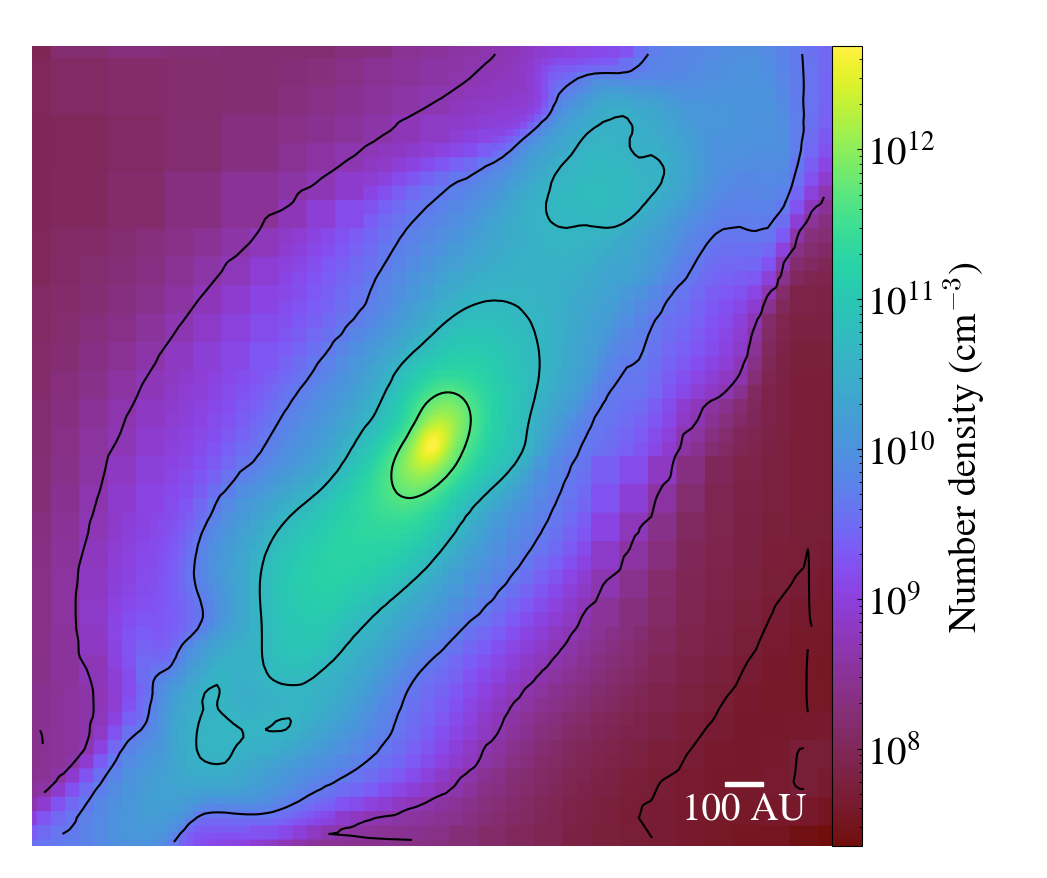}\par
    \includegraphics[width=0.3\textwidth]{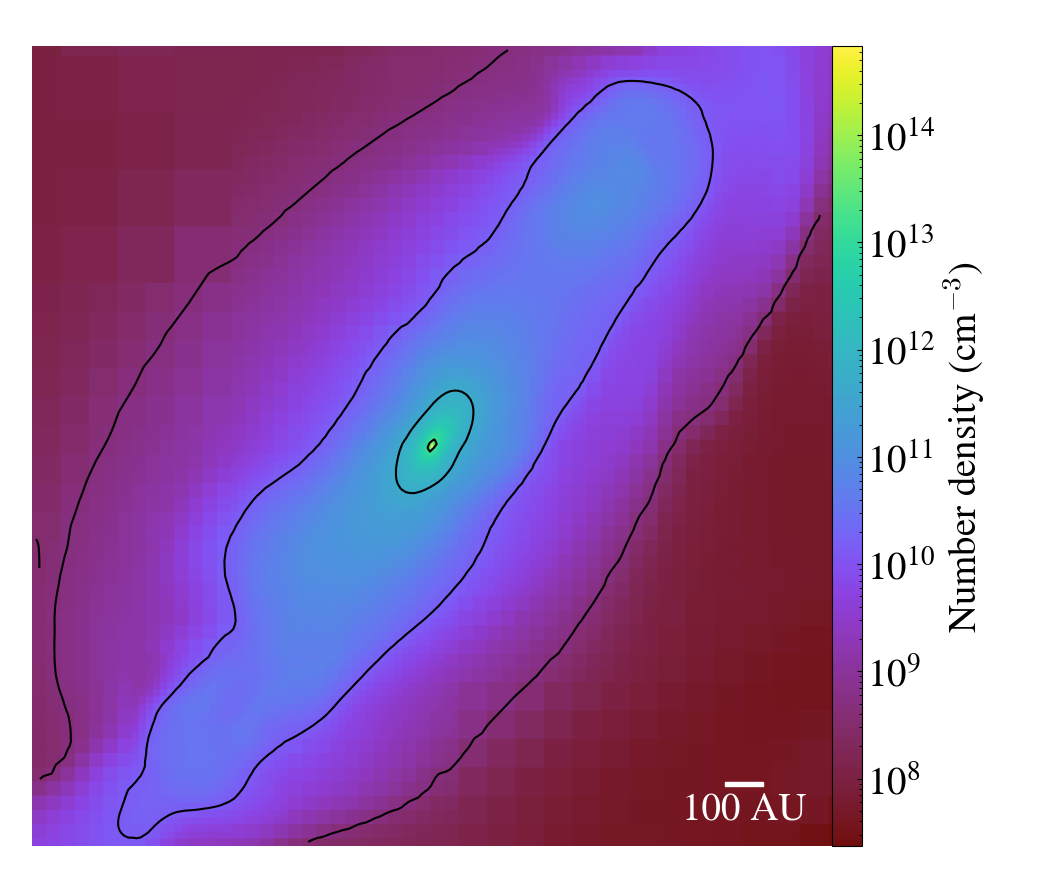}
\end{multicols}
\caption{Time evolution of the gas distribution around the central clump. {\it Top row:} HD case. {\it Bottom row:} Toroidal field for intermediate strength case (TI). Note that the spatial scale is larger for the panels in the first column, compared with the other two, which provide a zoom into the center. The first column displays the situation at maximum number density $10^{9} \pcc$, when the magnetic field is introduced, evolving via the MHD solver. In both cases, one central disc emerges, with slight differences in disc morphology caused by the magnetic field at this point. The second column, at density $10^{12} \pcc$, shows a clear break of the disc's radial symmetry in the TI case. The third column, at the final stage when the number density reaches $10^{15} \pcc$, shows the collapse into one central peak in both cases, with a difference in the gas distribution around it.}
\label{fig:timeevol}
\end{figure*}
\begin{figure*}
\begin{multicols}{3}
    \includegraphics[width=0.3\textwidth]{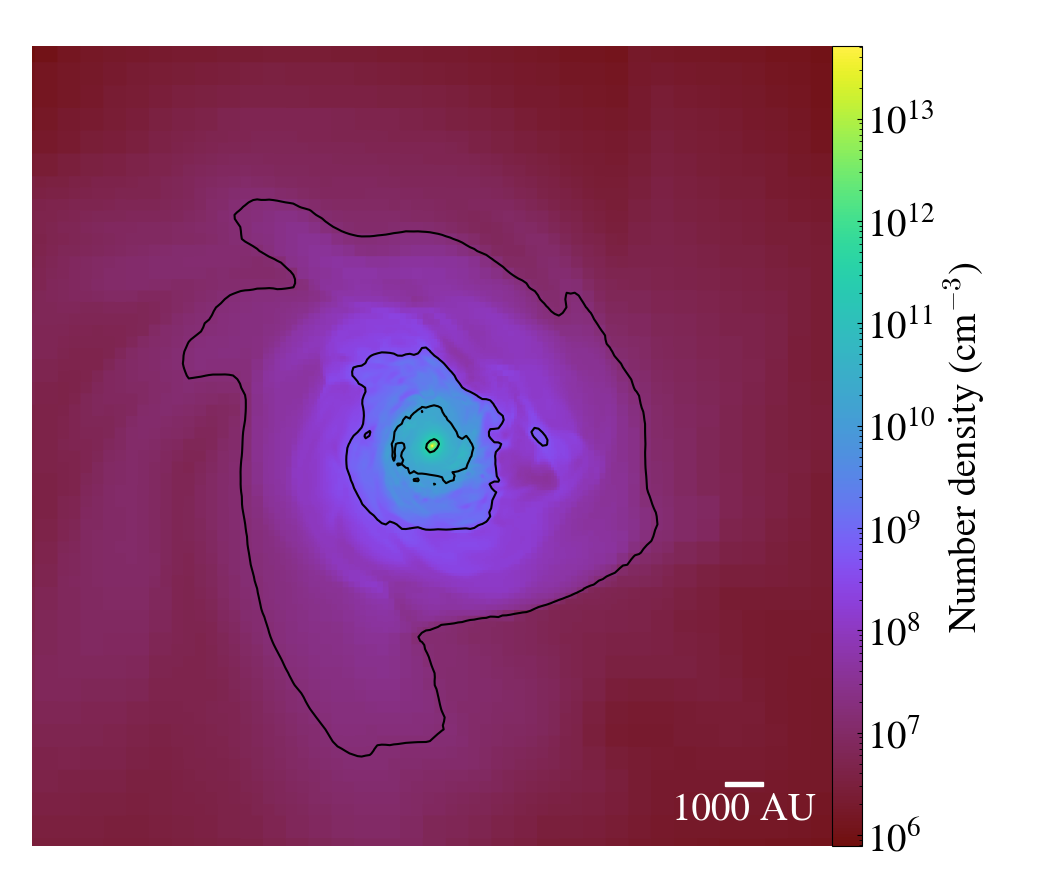}\par
    \includegraphics[width=0.3\textwidth]{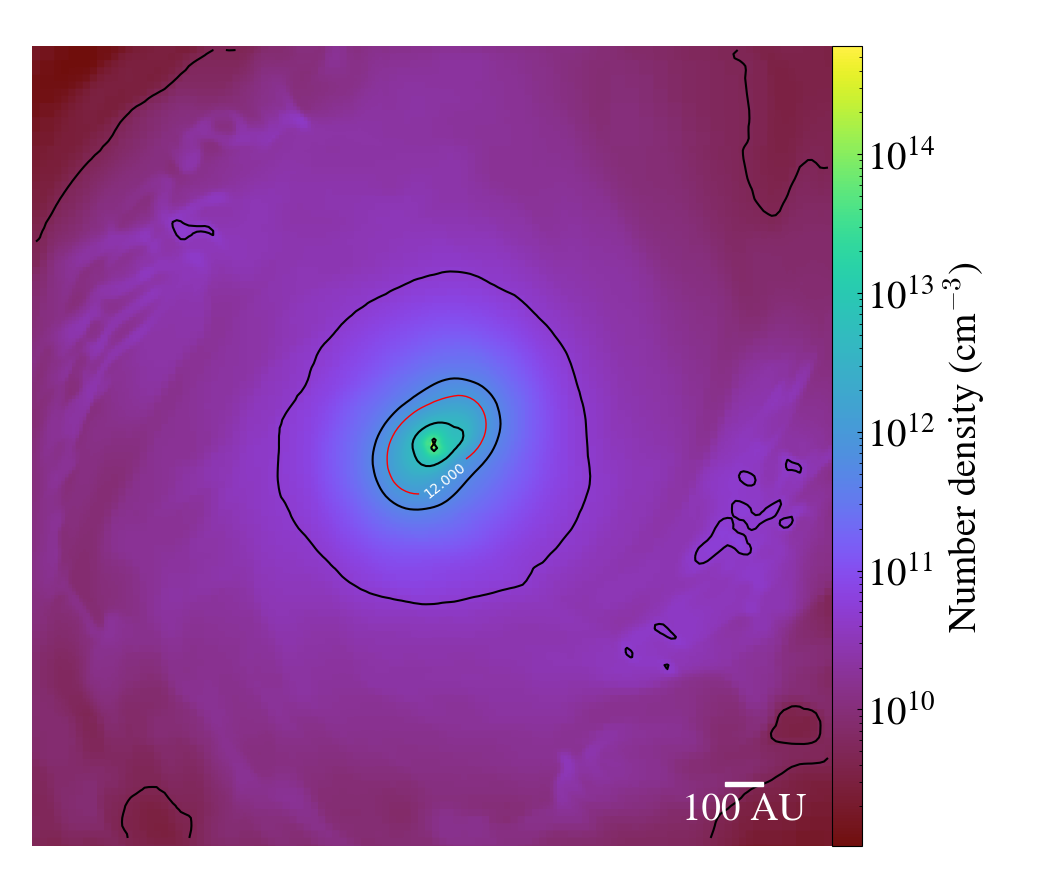}\par
    \includegraphics[width=0.3\textwidth]{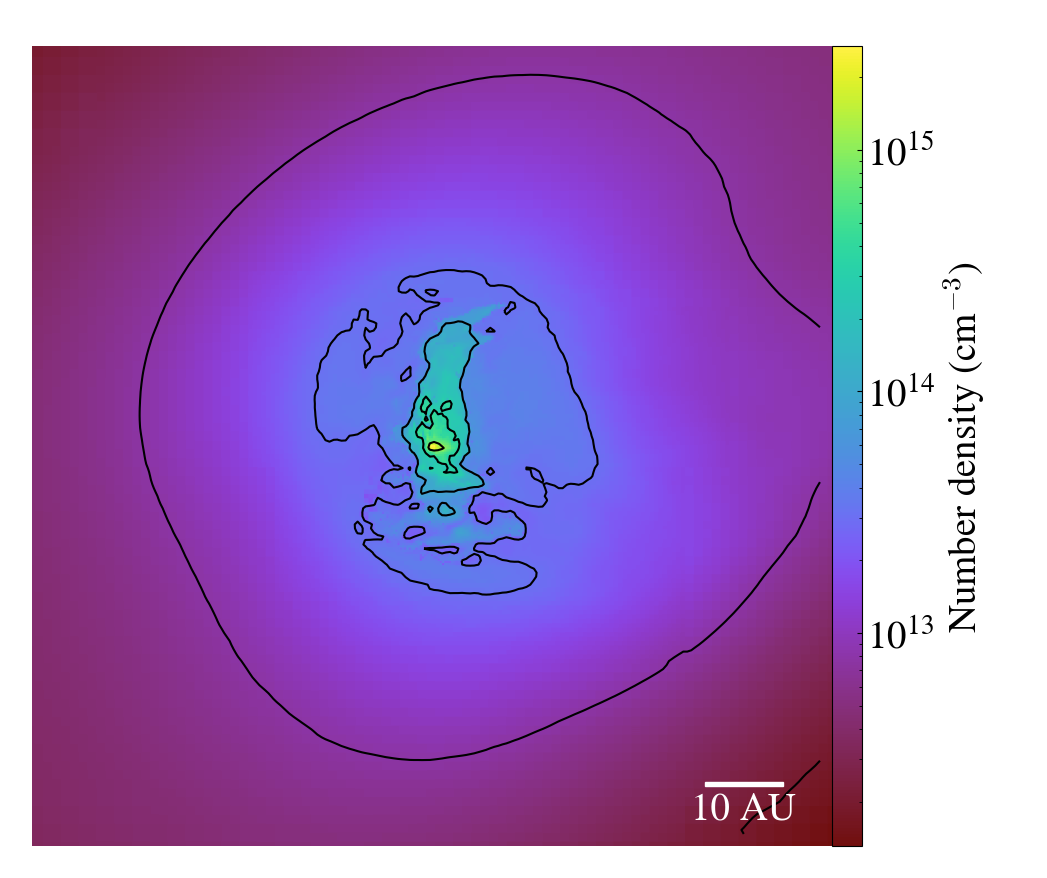}
\end{multicols}
\begin{multicols}{3}
    \includegraphics[width=0.3\textwidth]{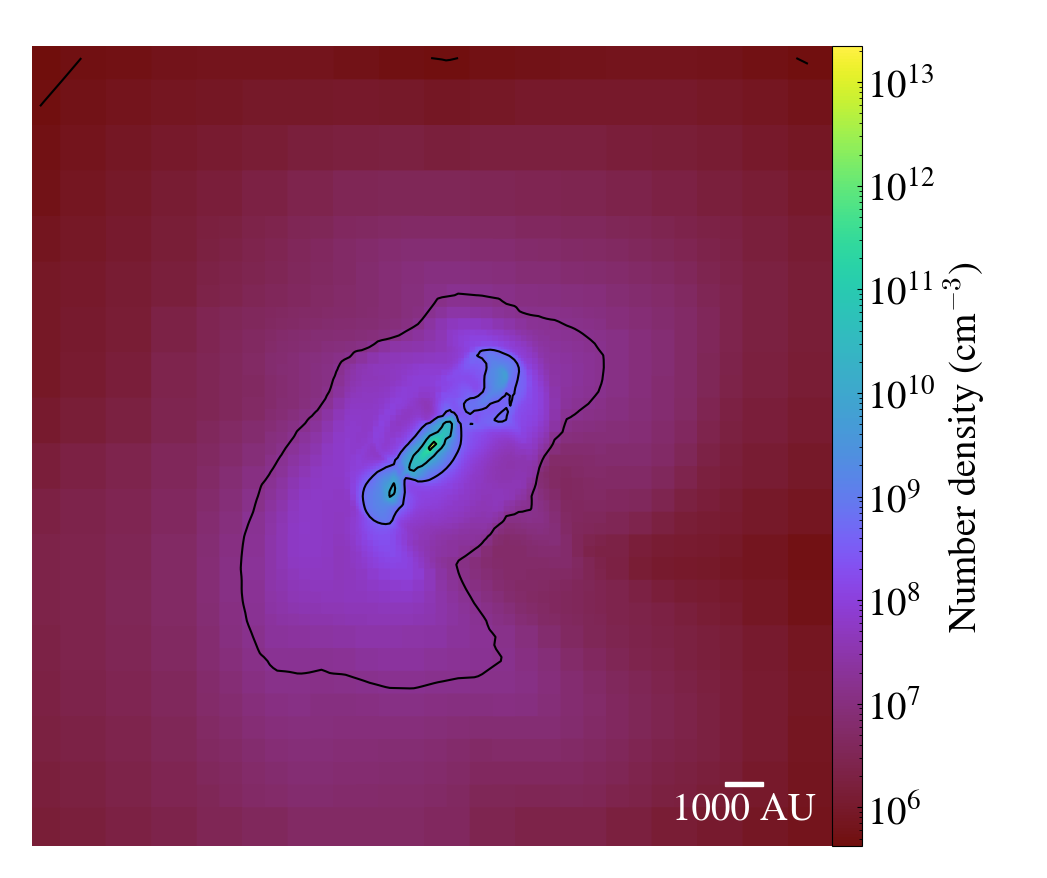}\par
    \includegraphics[width=0.3\textwidth]{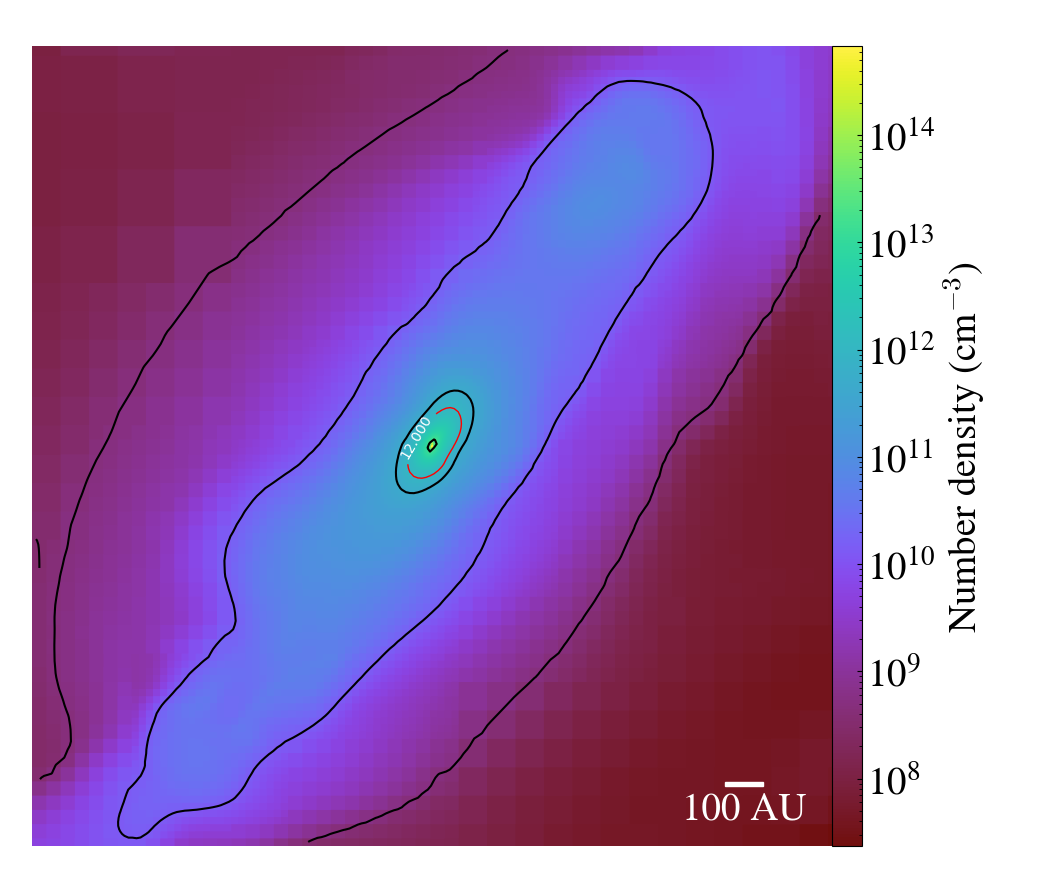}\par
    \includegraphics[width=0.3\textwidth]{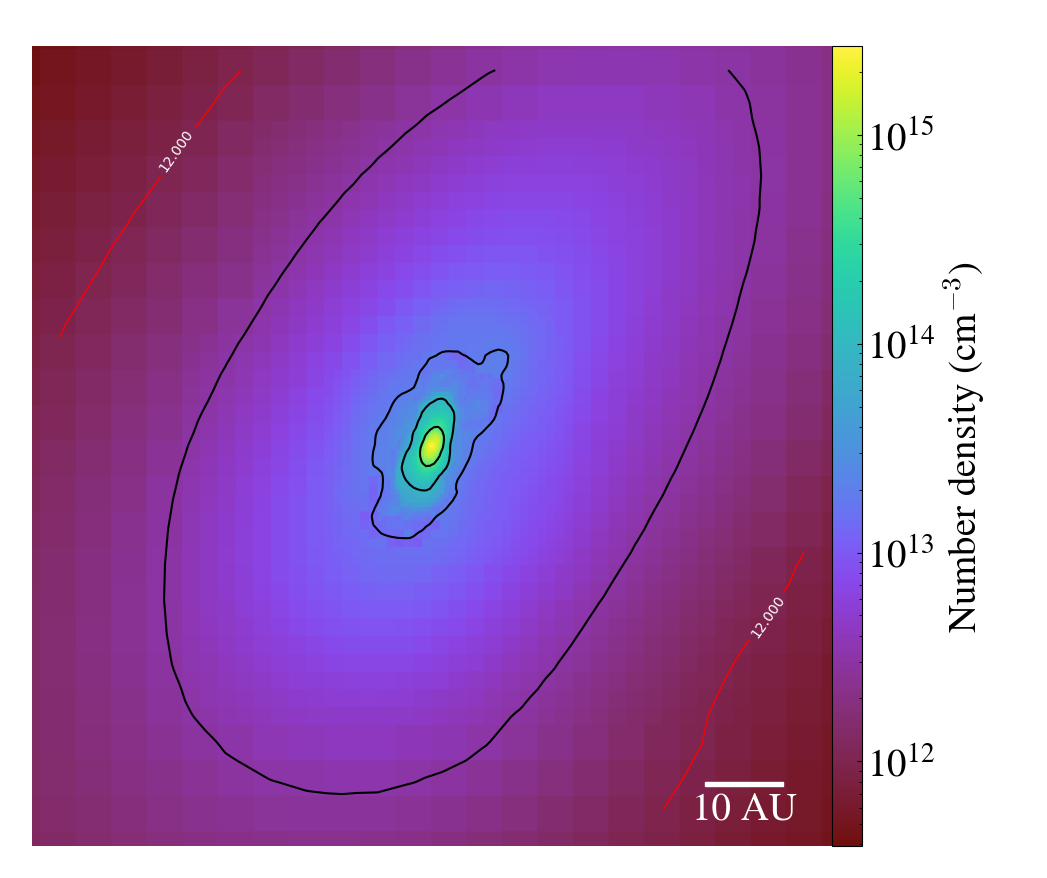}
\end{multicols}
\caption{Density distribution around the maximum density of $10^{15} \pcc$, to compare the final stage at different spatial scales. {\it Top row:} HD case. {\it Bottom row:} Toroidal field for intermediate magnitude (TI). Boxes of different lengths around the central peak are displayed: {\it Left column:} 0.1\,pc, showing the central disc. {\it Middle column:} $0.01$\,pc, showing the protostar, defined by its photospheric surface (red contour). {\it Right column:} $5\times 10^{-4}$\,pc, showing the situation inside the protostellar surface. This zoom-in indicates that the central peak in the HD case looks clumpy, in contrast to the TI case. However, such small scales are not reliably resolved here, and need to be investigated further in the high-resolution runs (see below).}
\label{fig:HDvsMHDI}
\end{figure*}

\begin{figure*}
\begin{multicols}{3}
    \includegraphics[width=0.3\textwidth]{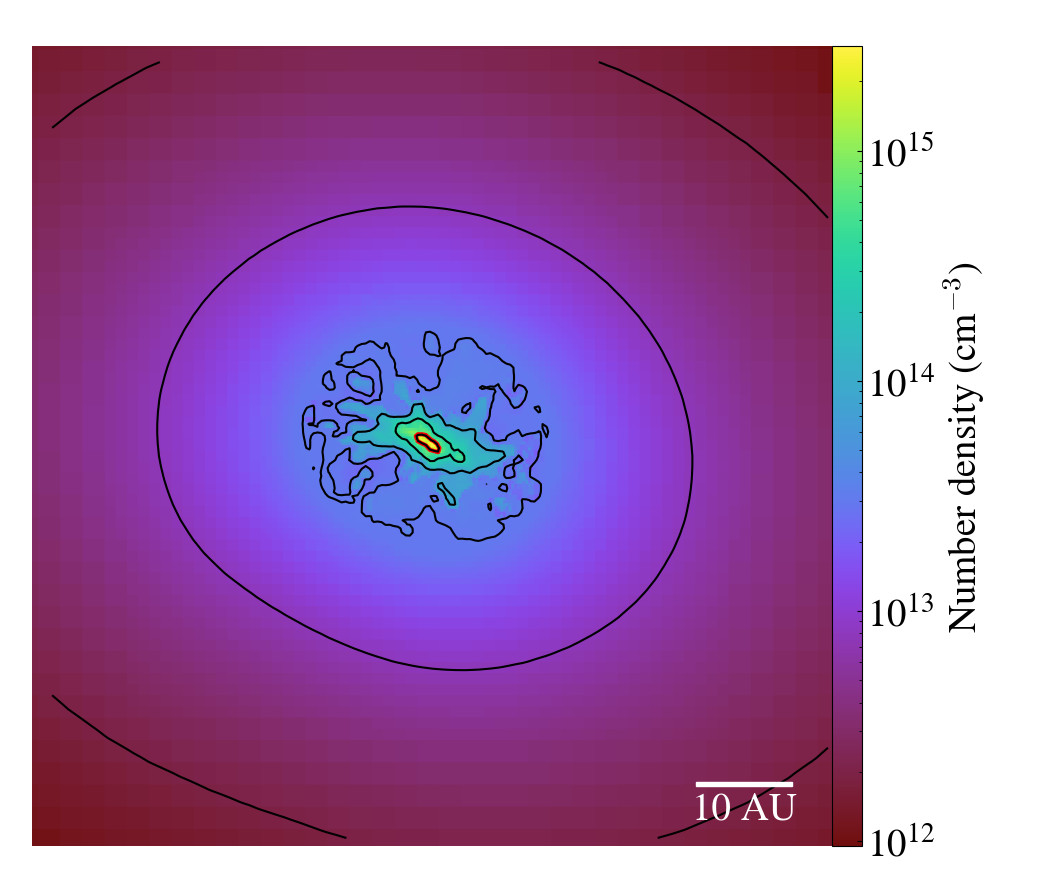}\par
    \includegraphics[width=0.3\textwidth]{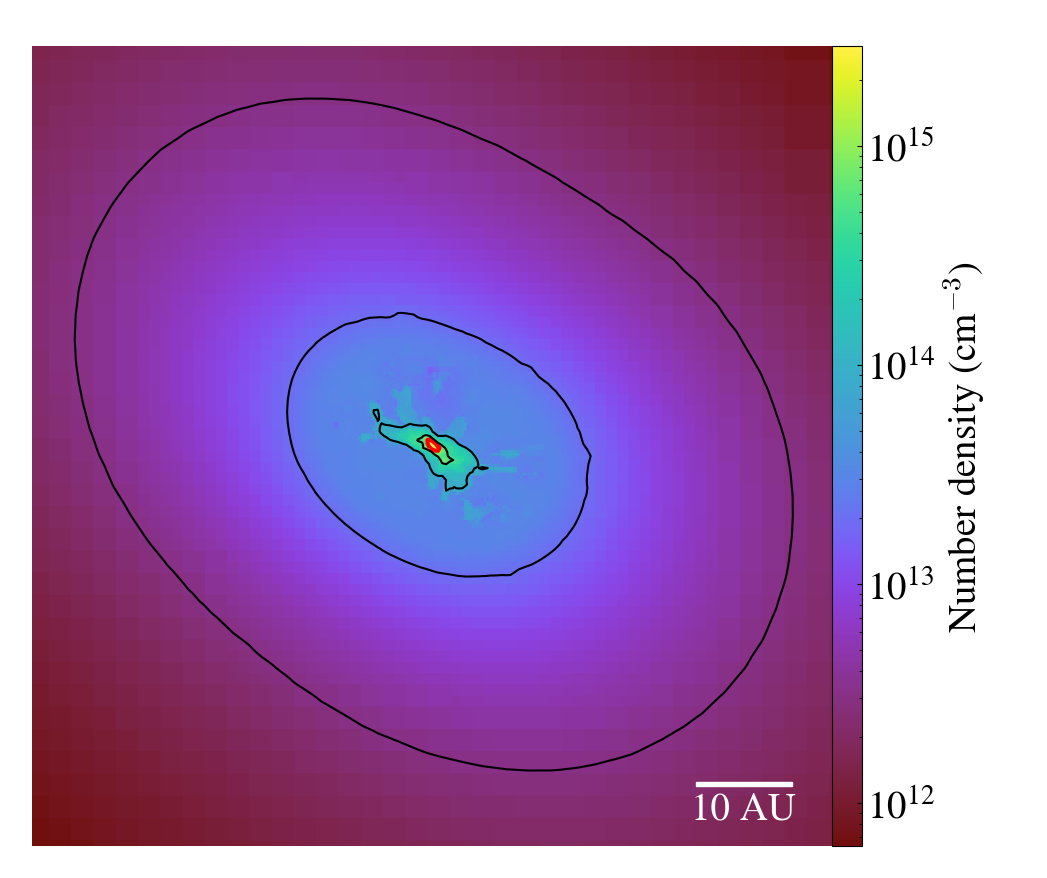}\par
    \includegraphics[width=0.3\textwidth]{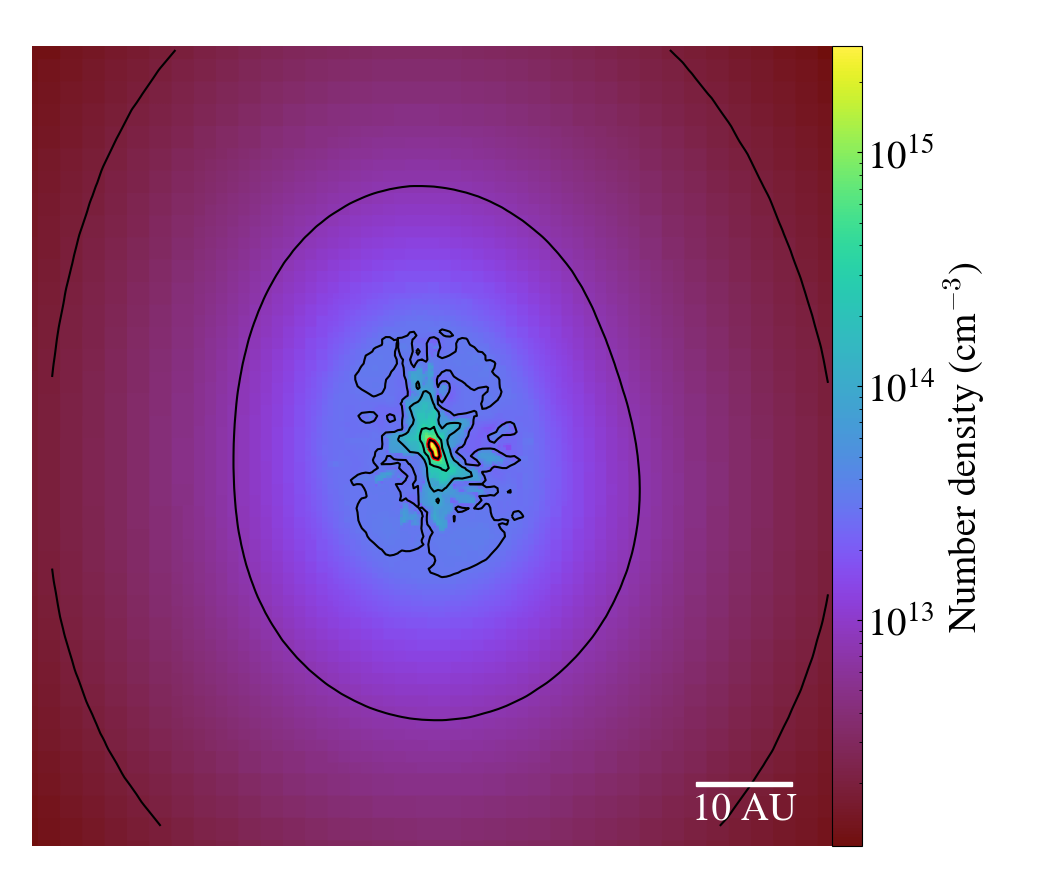}
\end{multicols}
\begin{multicols}{3}
    \includegraphics[width=0.3\textwidth]{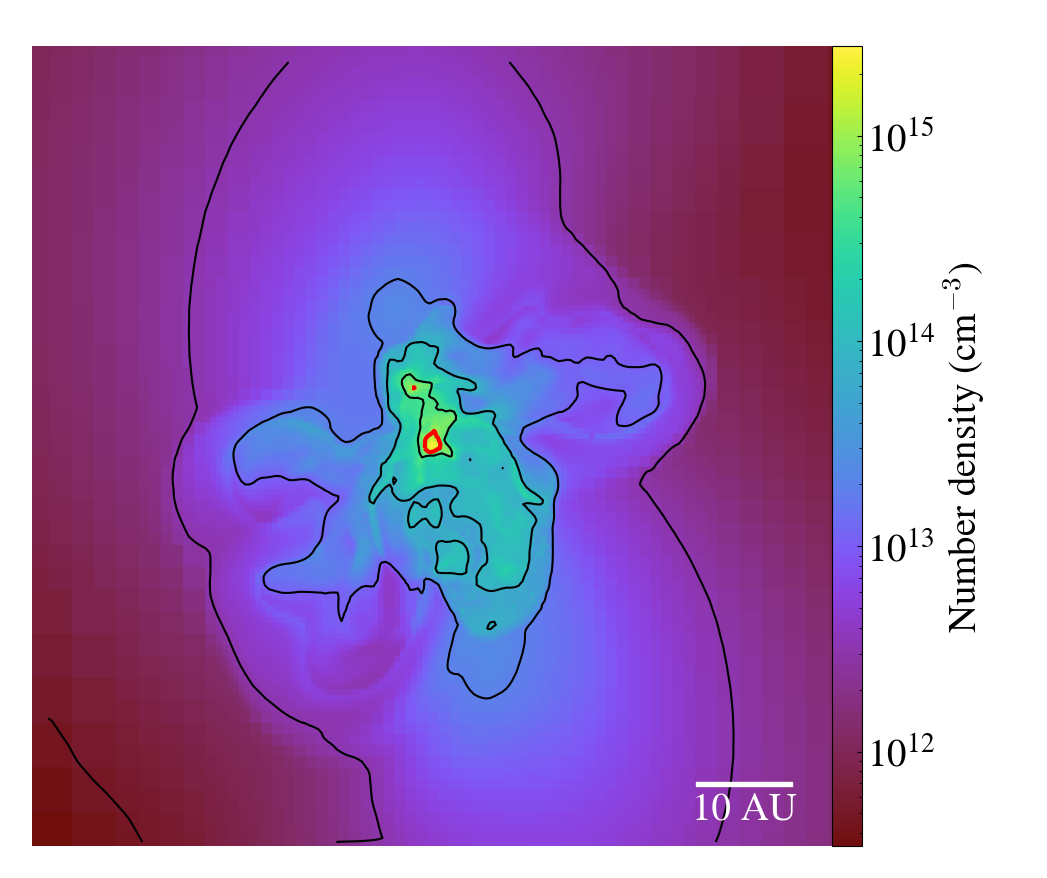}\par
    \includegraphics[width=0.3\textwidth]{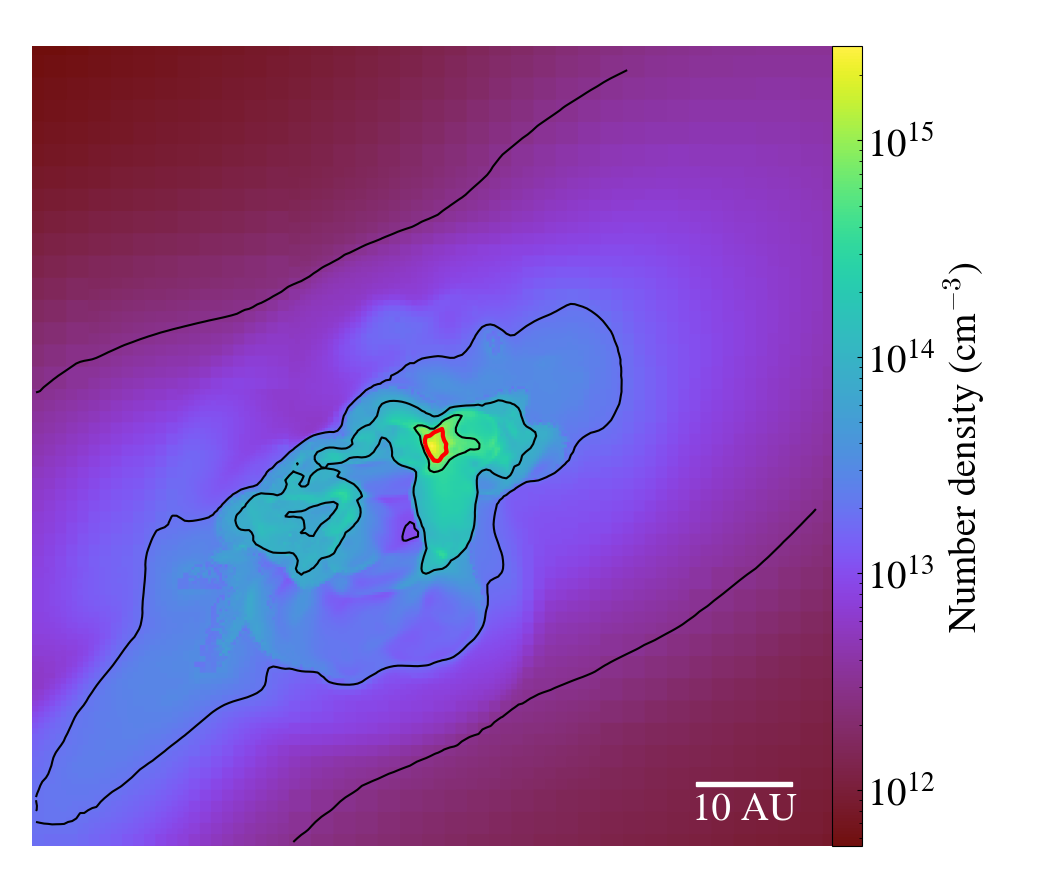}\par
    \includegraphics[width=0.3\textwidth]{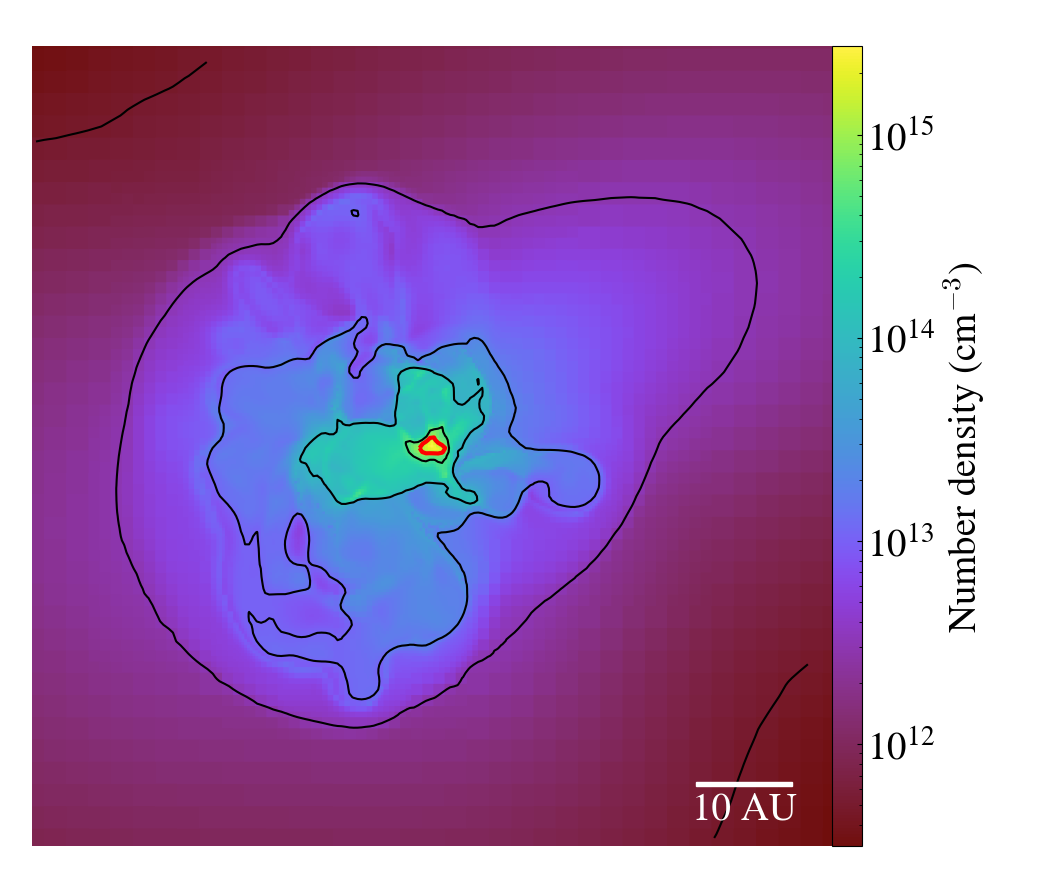}
\end{multicols}
\begin{multicols}{3}
    \includegraphics[width=0.3\textwidth]{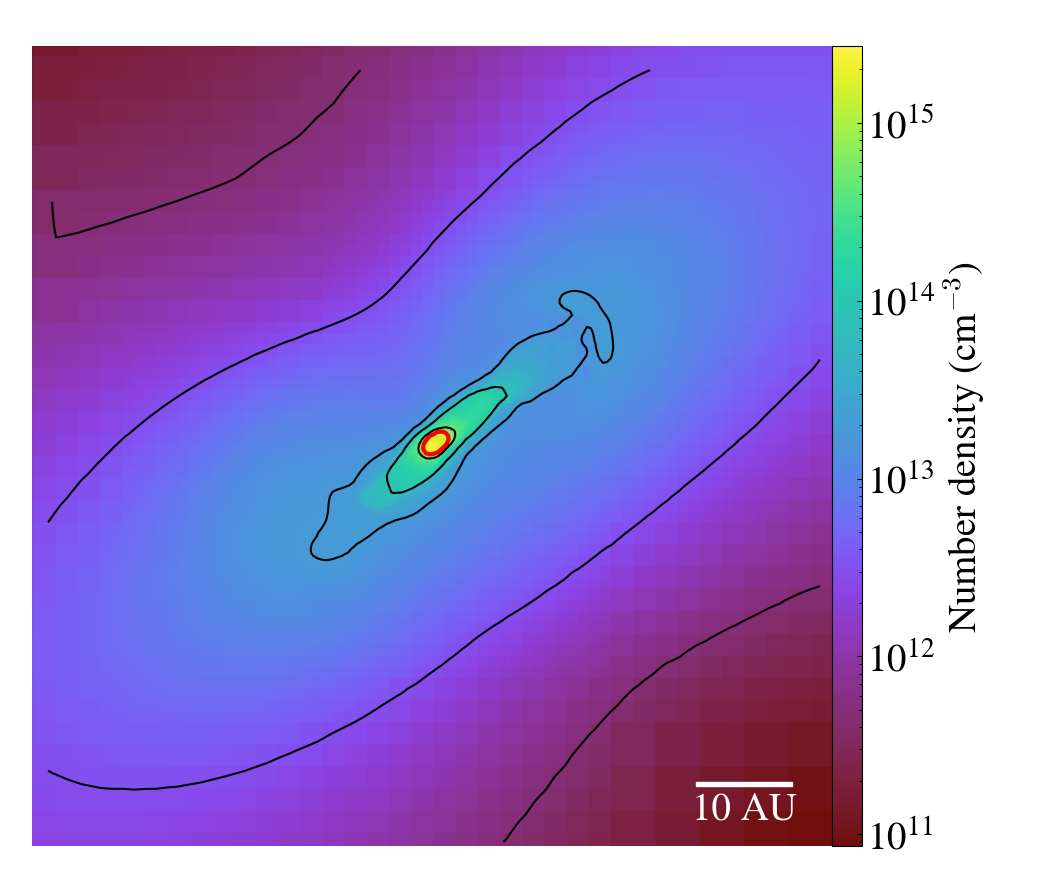}\par
    \includegraphics[width=0.3\textwidth]{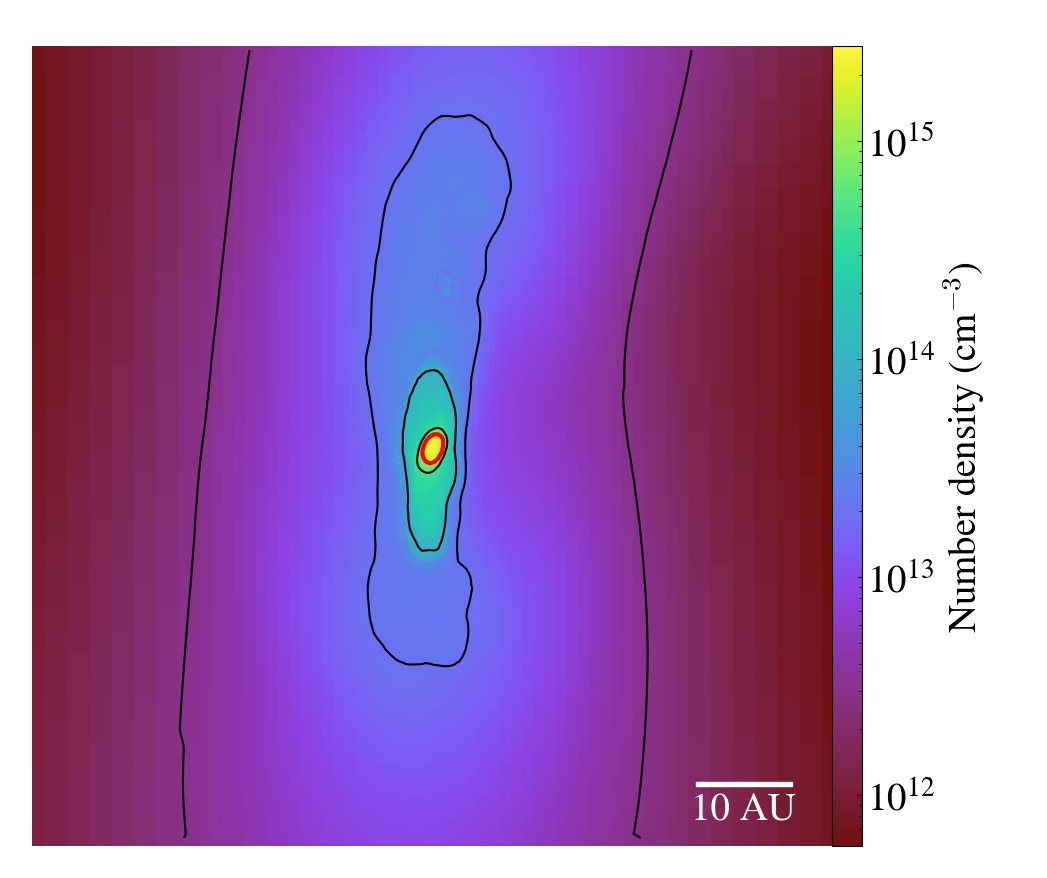}\par
    \includegraphics[width=0.3\textwidth]{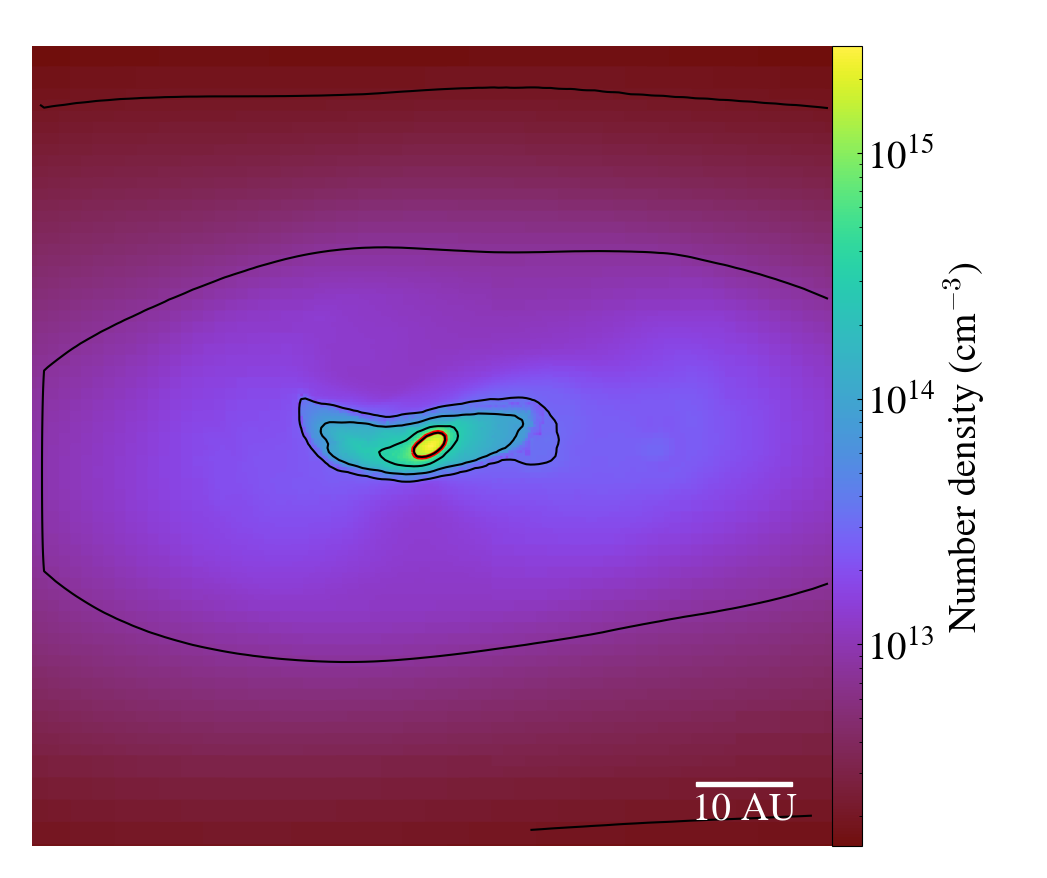}
\end{multicols}
\caption{Density distribution for the high-resolution runs, where the threshold number density for the artificial optical depth is $n_{\rm{th}}=10^{15}\pcc$. The protostellar contours are shown in red. {\it Top row:} HD case. {\it Middle row:} Intermediate-strength poloidal field case (PI). {\it Bottom row:} Intermediate-strength toroidal field case (TI). The columns represent, from left to right, projections along the $x$, $y$ and $z$ axes, respectively. We show the final evolutionary stage that we could reach with the available computational resources, after forming a protostar in each case. In the HD case, a clumpy disc is formed, demonstrating that the clumpiness exhibited in Fig.~\ref{fig:HDvsMHDI} is not a numerical artefact. The toroidal case forms a single protostar with a more compact disc, whereas in the PI case fragments are seen close to the central protostar.}
\label{fig:HDvsMHDI_noopacity}
\end{figure*}

\begin{figure*}
\begin{multicols}{2}
    \includegraphics[width=0.48\textwidth]{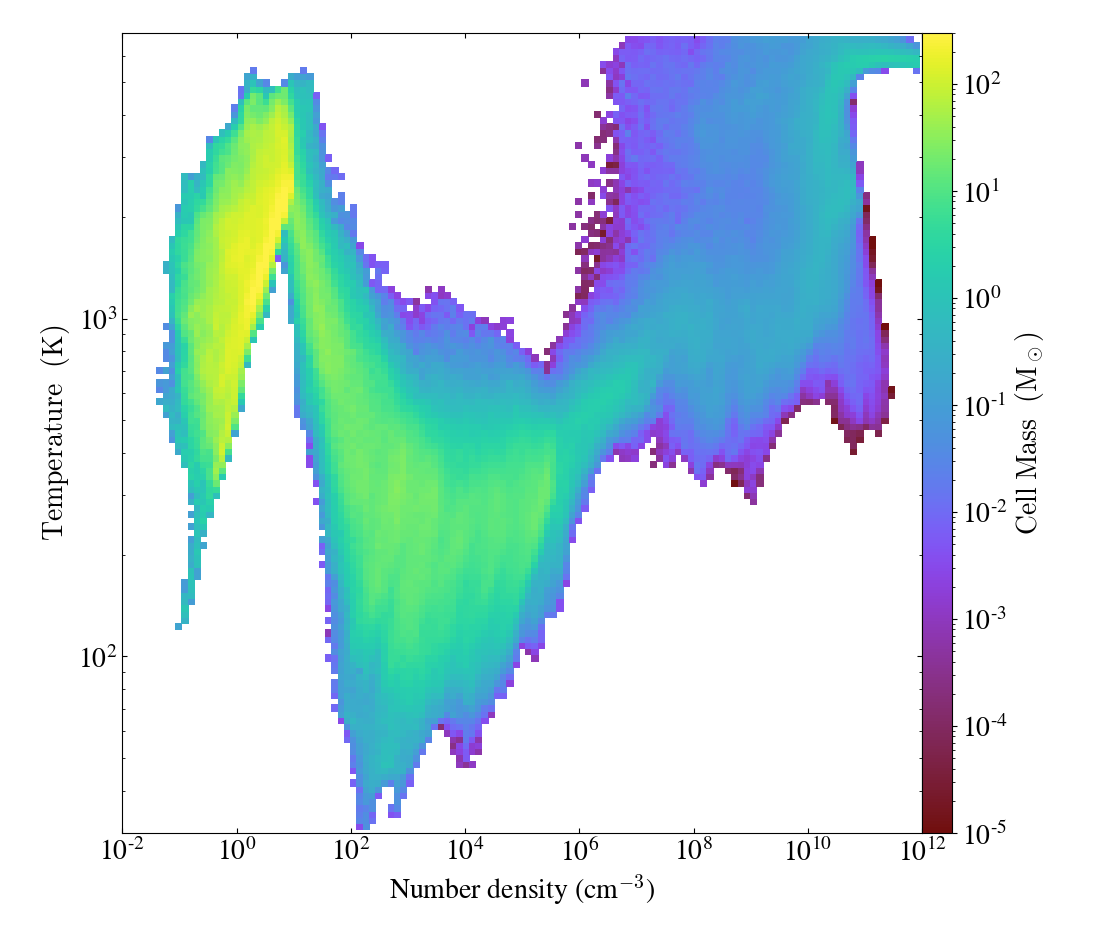}\par
    \includegraphics[width=0.48\textwidth]{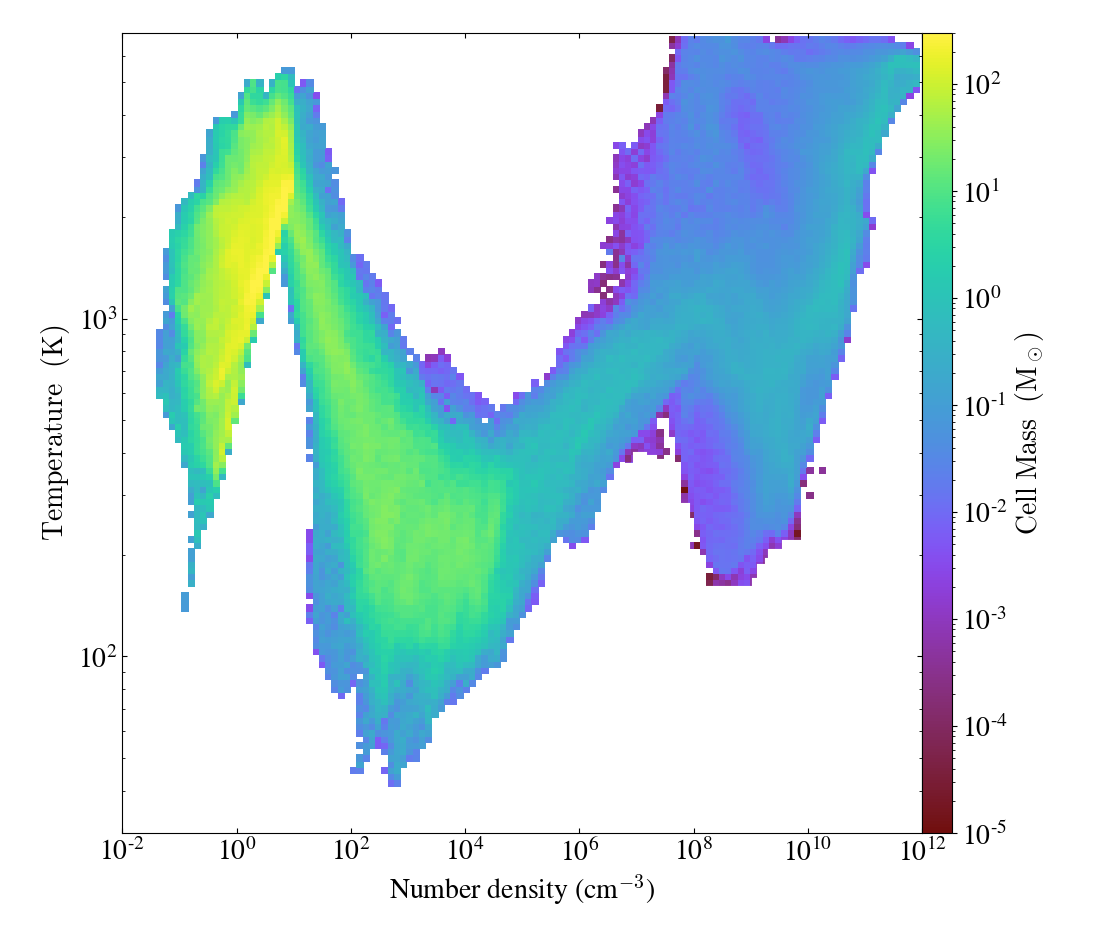}
\end{multicols}
\caption{Thermodynamical evolution in the collapsing primordial gas. We show the temperature vs. number density phase space at the final snapshots, below $n_{\rm th} = 10^{12}\pcc$. {\it Left panel:} HD case. {\it Right panel:} MHD case, specifically the TI run. The thermodynamical behaviour is overall very similar for both cases, but at $n\gtrsim 10^{8} \pcc$, temperatures are somewhat lower in the MHD case. This latter effect is caused by the slow-down of the compressional collapse in response to the additional magnetic pressure.}
\label{fig:comparisonTEMP}
\end{figure*}
\begin{figure}
    \includegraphics[width=0.47\textwidth]{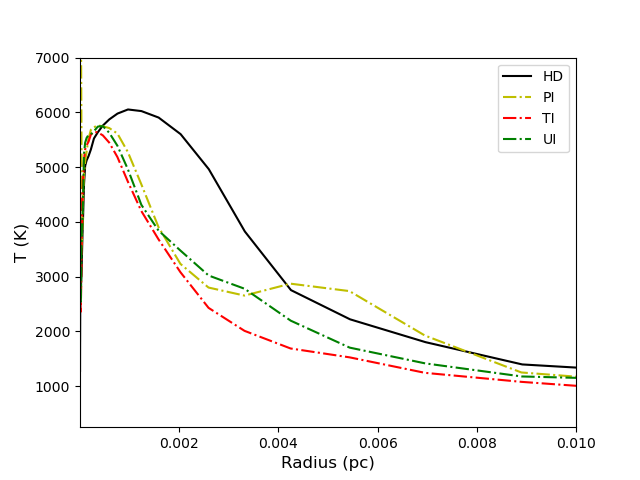}
    \caption{Temperature profile as a function of distance from the central peak at density $10^{15} \pcc$. Cases with different magnetic field geometries for intermediate initial strength are compared to the HD case, as indicated. As can be seen, the magnetic field leads to overall lower temperatures in the disc, due to the added magnetic pressure that delays the collapse. The central drop in temperature is a numerical artefact (see main text).}
    \label{fig:comparisonT}
\end{figure}

\begin{figure*}
\begin{multicols}{2}
    \includegraphics[width=0.48\textwidth]{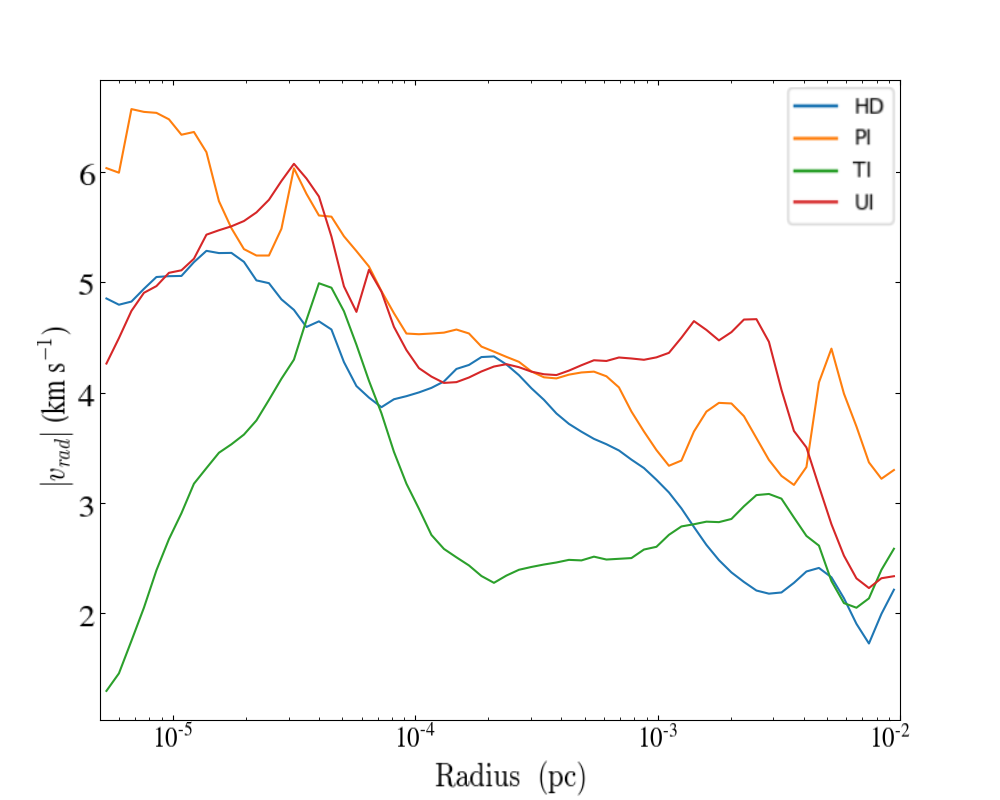}\par
    \includegraphics[width=0.48\textwidth]{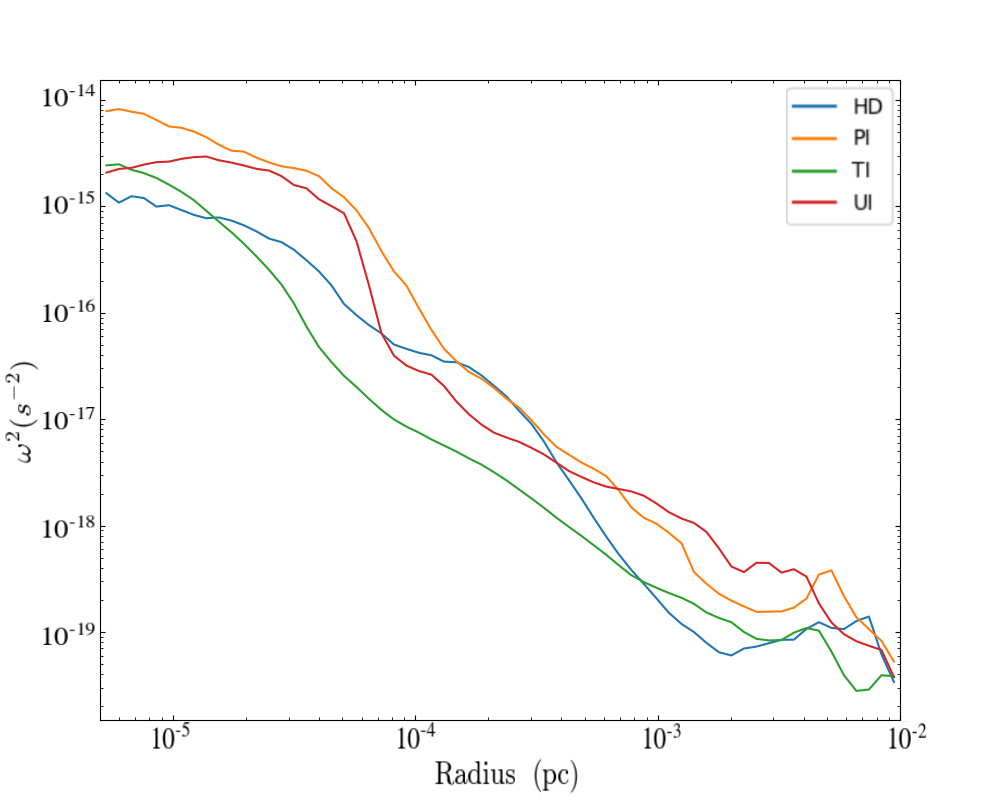}
\end{multicols}
\begin{multicols}{2}
    \includegraphics[width=0.48\textwidth]{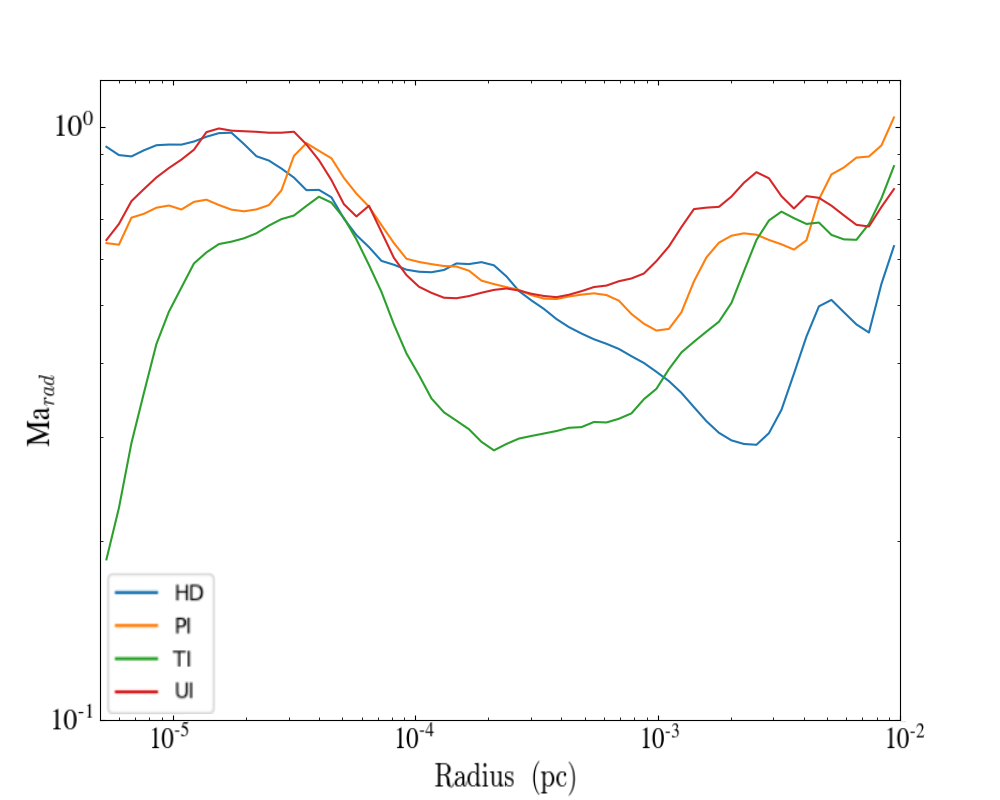}\par
    \includegraphics[width=0.48\textwidth]{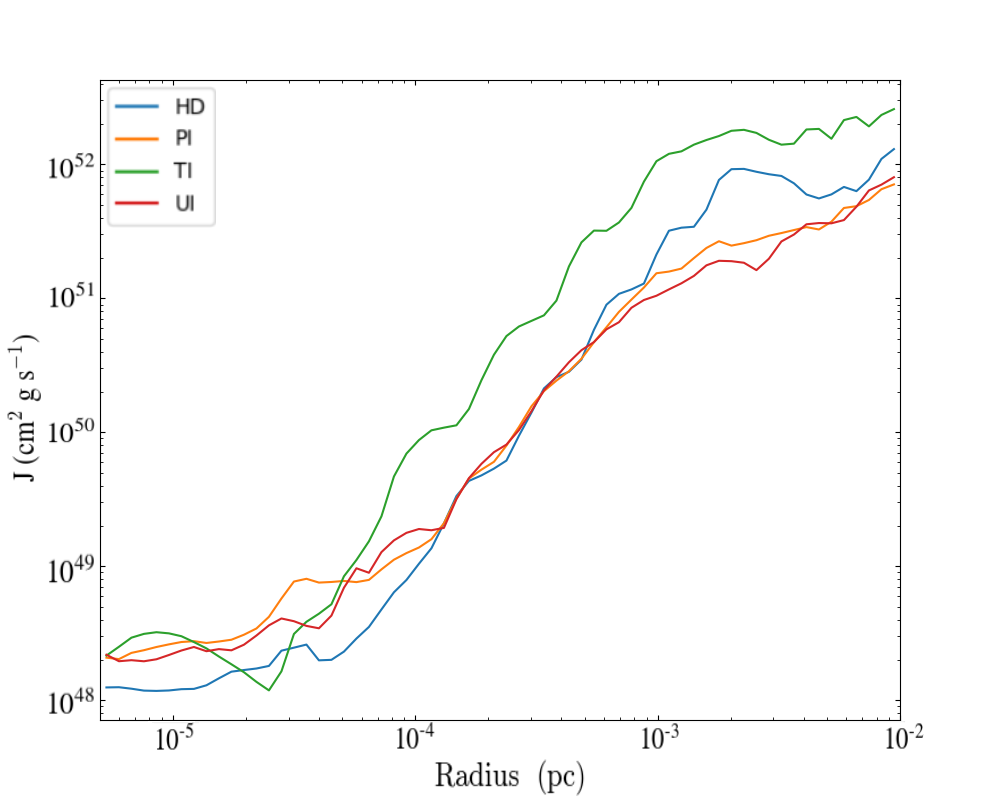}
\end{multicols}
\caption{Kinematic structure of the central star forming region. From top left to bottom right: radial velocity, vorticity squared, radial Mach number and magnitude of angular momentum vs. distance to the central peak with a density of $10^{15} \pcc$. Cases with different magnetic field geometries for intermediate-strength normalization are compared to the HD case. The magnetic field is slowing the collapse. The high vorticity region around the central peak indicates an overall increase in turbulent energy, whose detailed properties are not fully resolved here. As indicated by the radial Mach number profile, the gas in the disc is mostly subsonic. Considering the angular momentum profile, it is evident that the MHD (TI) case is particularly efficient in transporting angular momentum outwards.}
\label{fig:comparison2}
\end{figure*}
\begin{figure*}
\begin{multicols}{2}
    \includegraphics[width=0.48\textwidth]{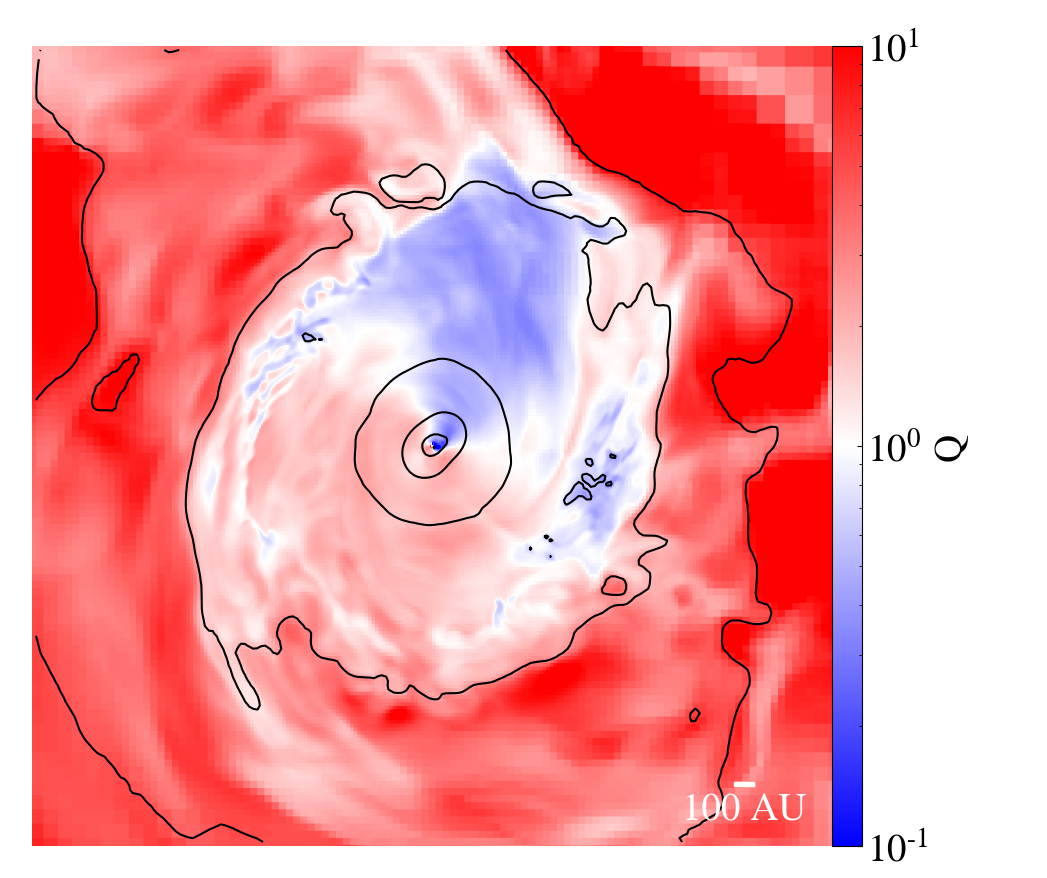}\par
    \includegraphics[width=0.48\textwidth]{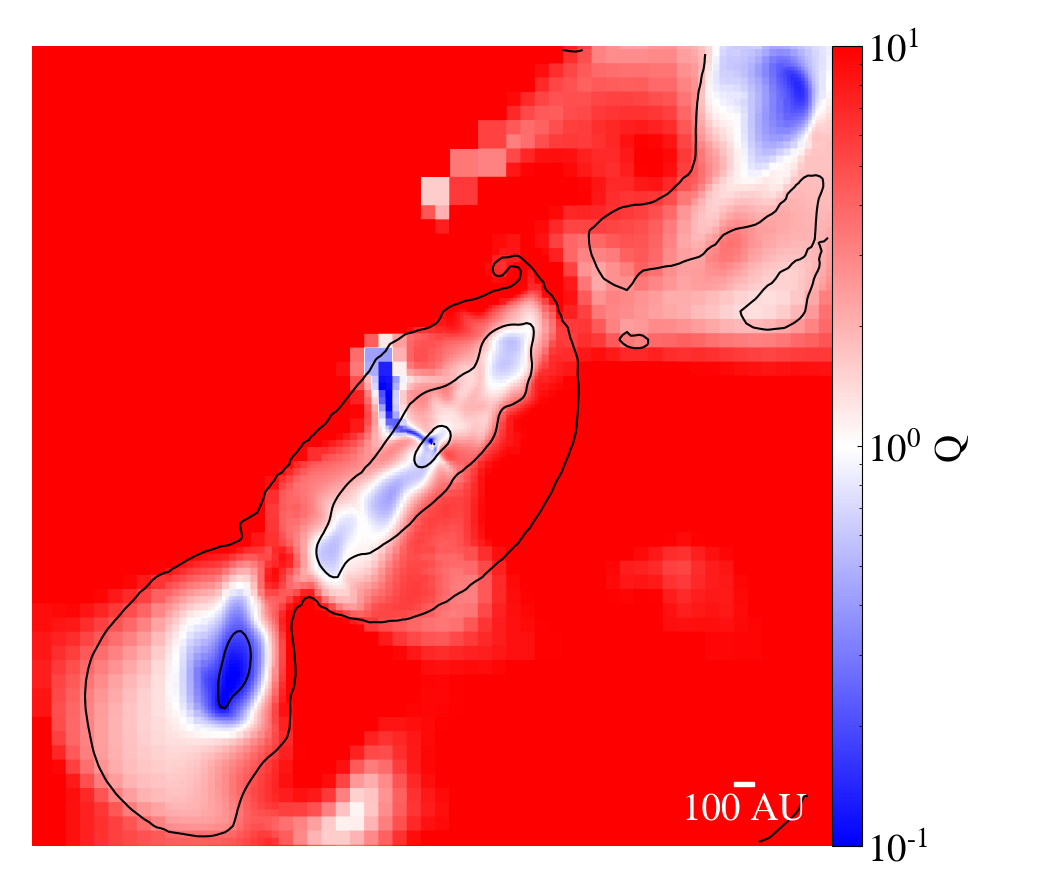}
\end{multicols}
\caption{Cross-sectional view around the central peak, illustrating the local Toomre stability criterion, $Q_\text{local}$, at the final time when the maximal density is $10^{15} \pcc$. {\it Left panel:} HD case. {\it Right panel:} Intermediate-strength toroidal field case (TI). In each panel, density contours are indicated with the black lines. In the HD case, the disc is axisymmetric, with several small clumps present, whereas the morphology in the MHD (TI) case is elongated, with two dominant sub-clumps.}
    \label{fig:toomreslices}
\end{figure*}
\begin{figure}
  \includegraphics[width=1\columnwidth]{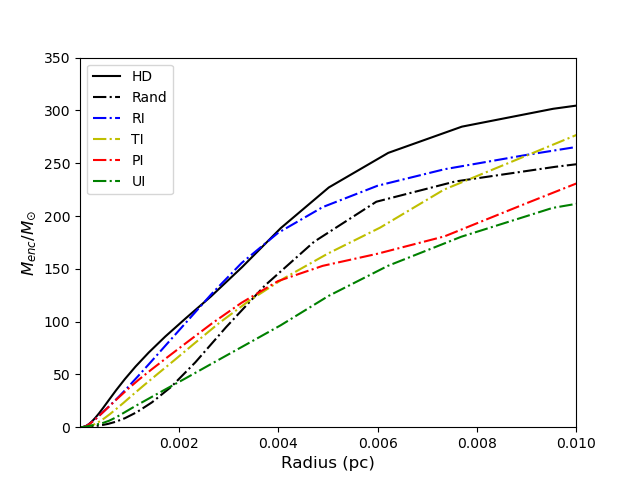}
  \caption{Enclosed mass in a sphere of radius $10^{-2}$~pc around the central density peak, as a function of radius for the HD and select MHD cases, specifically the Random, RI, TI, PI and UI configurations, at maximum number density reaches $10^{12} \pcc$. Very similar mass distributions pertain to the HD and MHD (RI) cases, due to the almost radial collapse in both cases. The magnetic field distorts the disc in the other cases, however, affecting the distribution of the enclosed mass.}
\label{fig:comparisonMASS}
\end{figure}

\begin{figure*}
\begin{multicols}{2}
    \includegraphics[width=0.47\textwidth]{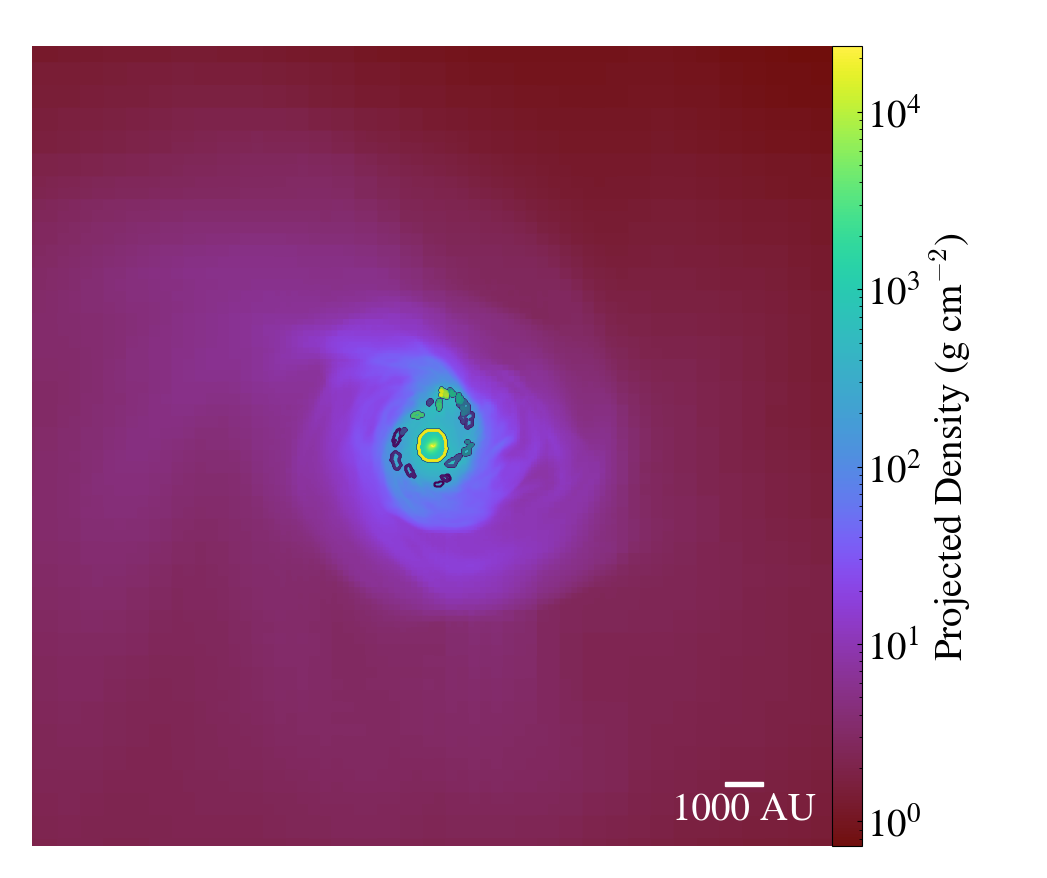}\par
    \includegraphics[width=0.47\textwidth]{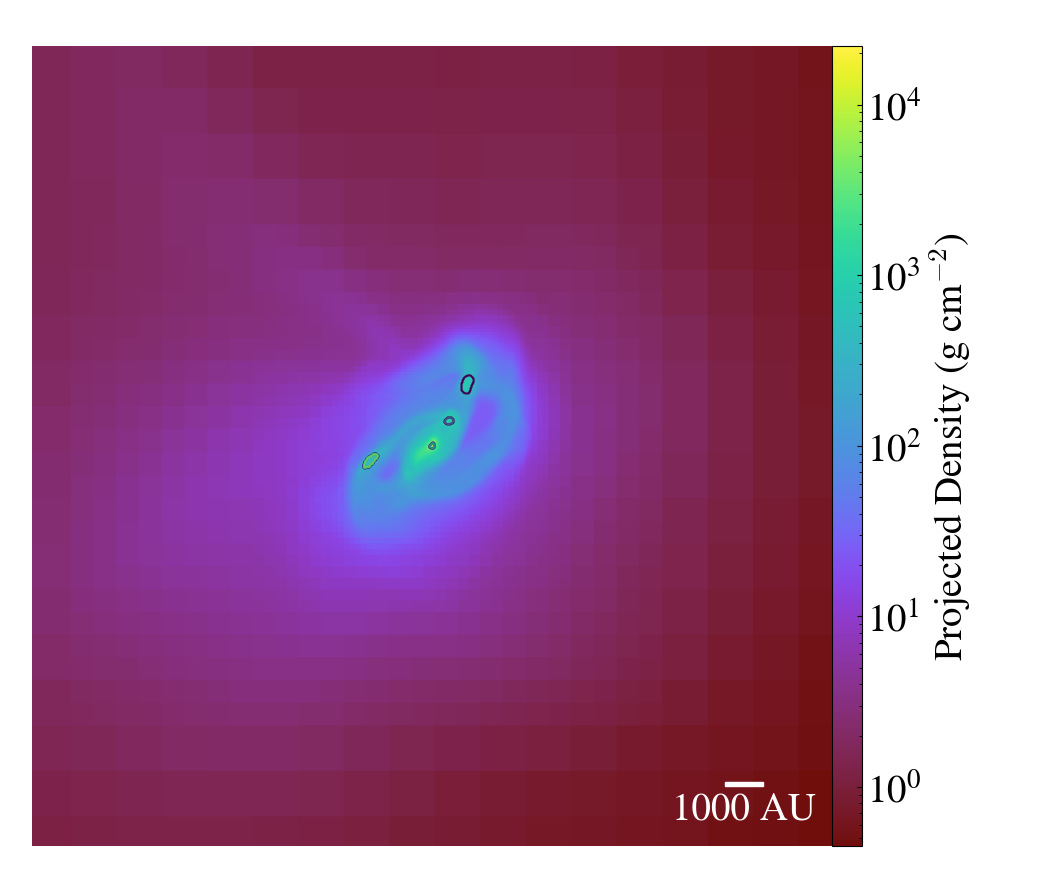}
\end{multicols}
\caption{Central density projection showing the distribution of clumps. {\it Left panel:} HD case. {\it Right panel:} Intermediate-strength toroidal field case (TI). A dominant central clump is surrounded by $\sim 50$ other low-mass fragments in the HD case, compared to a central clump with three smaller clumps in the MHD (TI) case.}
\label{fig:clumps}
\end{figure*}

\subsection{Initial conditions}
\label{sect2.3}
Our simulations start from cosmological initial conditions, calibrated with high-precision CMB observations (e.g. \citealt{planck15}). Specifically, we employ cosmological parameters as follows: a total matter density (dark matter and baryons) of $\Omega_m = 0.2603$, dark energy density $\Omega_{\Lambda} = 0.6911$, baryon density $\Omega_b = 0.0486$, and Hubble constant $h=0.6774$. The power spectrum, given by \citet{eisenstein1999}, assumes a spectral index of $n=0.961$, and is normalized to $\sigma_8 = 1.2$, which is artificially increased to compensate for the relatively small box size, effectively accelerating structure formation that might otherwise have been missed \citep[see table 1 in][]{greif2012}. The simulations are initialized at $z=100$ in a cosmological cube of comoving size $0.2 {\rm \,Mpc}/h$, with a base resolution of $256^3$, using the initial conditions generator MUSIC (e.g. \citealt{music}). The computational box has one nested grid centered on the most massive halo. 

The grid is refined on baryon and dark matter overdensities $3\times 2^{-0.2l}$, where $l$ is the AMR level, similar to \citet{wise2012,koh2016}. The exponent is chosen so that the refinement becomes super-Lagrangian, and the cells are refined more aggressively (e.g. \citealt{enzo}). The local Jeans length is resolved by 64 cells to avoid artificial fragmentation during gaseous collapse (e.g. \citealt{truelove1997}). 
If any of these criteria are met in a single cell, it is flagged for further spatial refinement. The simulations are stopped when the maximum number density reaches $10^{15} \pcc$, at maximum level of refinement 25, equivalent to a spatial resolution of $1\times10^{-6}\,{\rm pc}/h$ in the comoving frame. 

\begin{figure*}
\begin{multicols}{2}    
   \includegraphics[width=0.48\textwidth]{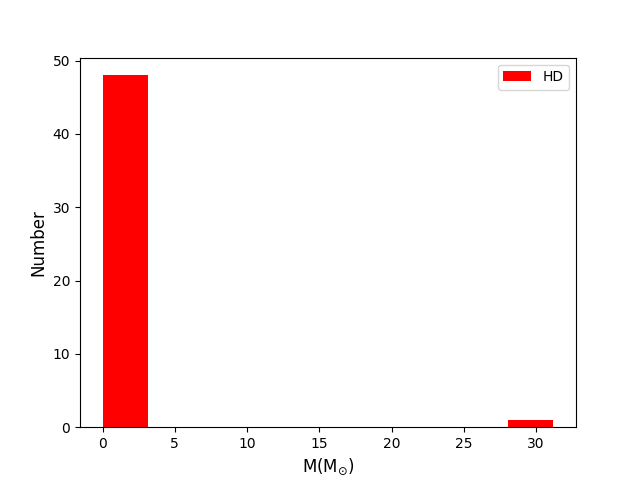}\par
    \includegraphics[width=0.48\textwidth]{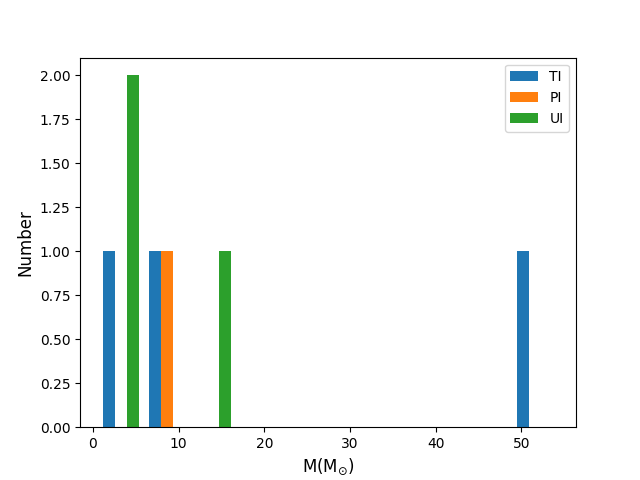}
\end{multicols} 
\caption{Distribution of clump masses, evaluated when a maximum density of $10^{15} \pcc$ is reached. {\it Left panel:} HD case. {\it Right panel:} MHD cases with intermediate-strength normalization. The left panel shows that in the HD case, a primary clump is formed of mass $\sim30\,M_{\odot}$, together with several clumps of masses $\sim\,1 M_{\odot}$, close to the Jeans mass evaluated at $n_\text{th}=10^{12} \pcc$. These low-mass fragments can be seen in the top row of Fig.~\ref{fig:clumps}. In contrast, for the MHD cases, only a few clumps are formed, with somewhat larger fragment masses, above the Jeans mass at the resolution limit. This re-emphasizes the overall conclusion that the magnetic field inhibits the small-scale fragmentation in the disc.}
\label{fig:hist}
\end{figure*}

\subsection{Incorporating MHD effects}
\label{sect2.4}
\subsubsection{Methodology}
\label{sect2.4.1}
For each grid cell, \texttt{Enzo} solves the equations describing the internal gas energy and the total energy, as a function of time. This dual energy formalism ensures that the method yields the correct entropy jump at strong shocks, and also delivers accurate pressures and temperatures in cosmological hypersonic flows. For the runs without magnetic fields (hereafter termed hydro runs), the Piecewise Parabolic Method (PPM) is used (e.g. \citealt{piecewise}). On the other hand, to solve the cosmological MHD equations (e.g. \citealt{enzo}), the Harten-Lax-van~Leer (HLL) Riemann solver is employed (e.g. \citealt{toro2013}). The solenoidal constraint $\bm{\nabla} \cdot \bm{B} = 0$ is enforced according to the Dedner scheme (e.g. \citealt{dedner}).

Since we here study the formation of the first stars in the Universe, the Lyman-Werner (LW) radiation background is not included in the present simulations. The LW photons with energies just below the H-ionizing threshold would otherwise act to photo-dissociate H$_{\rm 2}$, the main cooling agent in primordial gas, at subsequent stages of cosmic star formation \citep[e.g.][]{Safranek2012}. Therefore, H$_{\rm 2}$ self-shielding does not need to be included as well.

When performing our MHD calculations, the amplification of the seed magnetic field to its fully-developed asymptotic strength is not treated in a self-consistent way. This has been attempted in \cite{turk2012}, who found that the amplification process cannot be resolved accurately, due to the extreme dynamic range of the turbulent cascade involved. We therefore carry out a suite of numerical experiments, where we {\it assume} that small-scale turbulent dynamos have amplified the fields to close to equipartition with the turbulent kinetic energy \citep[e.g.][]{sur2010, schober2012}. We also consider cases where subsequent large-scale kinematic dynamo action has established a large-scale, ordered field configuration \citep[e.g.][]{tan2004}. For simplicity, we thus insert an already fully-developed magnetic field into the simulation box, if the number density exceeds $10^8\pcc$, with an amplitude to be discussed next. 

For comparison, we also carry out MHD calculations with a uniform magnetic field initialized at the beginning of the simulation at redshift $z=100$, similar to \citet{turk2012}. We consider seed field strengths of $10^{-12}$, $10^{-10}$ and $10^{-8}$\, G. Subsequently, the magnetic field is evolved self-consistently via the MHD solver described above.

\subsubsection{Field amplitude}
\label{sect2.4.2}
What are expected field strengths in Pop~III star forming regions? In surveying relevant results, according to \cite{xu2008}, the Biermann battery in conjunction with compressional amplification can result in fields with strengths of $B = 10^{-9}$\,G at number density $n = 10^{10} \pcc$ at the center of a cosmological halo where a Pop~III star is expected to form. Hence, the magnetic fields created by the Biermann battery are dynamically unimportant at all densities below $n = 10^{10} \pcc$, where the ratio of thermal gas to magnetic pressure is $\beta= P_{\rm th}/P_{\rm B} \geq 10^{15}$ at all times during the evolution of the minihalo. 

Significantly larger field strengths are reached when dynamo activity is considered, however. In fact, \cite{sharda2021} showed that the small-scale dynamo increases the turbulent magnetic field to the level of 1 to 10 percent of equipartition, in agreement with \cite{federrath2011,federrath2014}.
\cite{schleicher2010} and \cite{mckee2020} also noted that once the field reaches equipartition, it will remain there as the collapse continues, so that the field will increase as $\rho ^{1/2}$ (for a constant turbulent velocity) rather than the $\rho ^{2/3}$ behaviour for compressional amlification under flux freezing. In \cite{stacy2022}, the field actually achieved equipartition at number density $10^{12} \pcc$, and half the equipartition value when the number density reaches $10^{8}\pcc$. The latter is the number density where magnetic fields are introduced in our simulations. 

When inserting the fields in our simulations, we specifically assume that the magnetic energy is nearly in equipartition with the thermal energy of the gas, such that \citep[e.g.][]{Hirano2018}:
\begin{equation}
\label{Eq.Bsat}
    \frac{B^2_{\rm sat}}{8 \pi}= \eta \ c_{\rm s}^2 \ \rho \mbox{\ ,}
\end{equation}
where $c_{\rm s}$ is the sound speed, $\rho$ the local gas density of the cell, and $\eta$ a free parameter between 0 and 1. We consider three values for the coupling efficiency, $\eta=0.1$, $0.5$ and $0.9$, to denote cases of weak, intermediate and strong fields, respectively. Hence, in our calculations, we assess the importance of MHD effects on Pop~III protostar formation under different field strengths.

\subsubsection{Field geometry}
\label{sect2.4.3}
In carrying out our suite of numerical experiments, we consider several illustrative configurations for the magnetic field geometry: uniform, radial, toroidal and poloidal, each with the three values of $\eta$ discussed above. In addition, we also consider a random field configuration, corresponding to a situation where the large-scale organization of the small-scale turbulent fields did not occur. For our idealized approach, at every timestep the magnetic field is added locally in cells, where the number density exceeds $10^{8} \pcc$, providing three-dimensional Cartesian components.

With the magnitude, $B=B_\text{sat}$, given by equation~(\ref{Eq.Bsat}), we can specify our select field configurations, projected onto the Cartesian grid. The uniform field has equal components along the $x, y {\rm,\, and\, } z$ axes, such that:
\begin{equation}
\label{Eq.Buniform}
    \bm{B_{\rm uniform}}=\frac{B}{\sqrt{3}}(\bm{\hat{i}} + \bm{\hat{j}} + \bm{\hat{k}}) \mbox{\ ,}
\end{equation}
where $\bm{\hat{i}}, \bm{\hat{j}} {\rm,\, and\, } \bm{\hat{k}}$ are the unit vectors in Cartesian coordinates. A radial field is also considered, as follows:
\begin{equation}
\label{Eq.Brad}
    \bm{B_{\rm rad}} = - B\left(\frac{x}{r}\ \bm{\hat{i}} + \frac{y}{r}\ \bm{\hat{j}} + \frac{y}{r}\ \bm{\hat{k}}\right) \mbox{\ ,}
\end{equation}
where coordinates are expressed with respect to the location of the maximum density, and $r=(x^2+y^2+z^2)^{1/2}$ is the radius. By construction, the position of the density maximum is excluded ($r \neq 0$), such that the field remains divergence-free. 

In order to define the toroidal and poloidal fields, we use spherical coordinates (r,$\theta$,$\phi$), with $\bm{\hat{r}}$, $\bm{\hat{\theta}}$ and $\bm{\hat{\phi}}$ representing the corresponding unit vectors. The polar and azimuthal angles are as usual given via:
\begin{equation} 
\begin{aligned}
    \theta &= \arccos{\frac{z}{r}}\mbox{\ ,}\\
    \phi &= \arctan{\frac{y}{x}}\mbox{\ .}
\end{aligned}
\end{equation}
The toroidal field has a component only along $\bm{\hat{\phi}}$, when considering the canonical toroidal-poloidal field decomposition, which again has to be projected onto the Cartesian system to be implemented in \texttt{Enzo}: 
\begin{equation}
\label{Eq.Bpol}
    \bm{B_{\rm tor}} = B\left(- \sin{\phi}\ \bm{\hat{i}} + \cos{\phi}\ \bm{\hat{j}}\right)\mbox{\ .}
\end{equation}
Similarly, the poloidal field has a component only along $\bm{\hat{\theta}}$, which can be written in Cartesian coordinates as:
\begin{equation}
\label{Eq.Btor}
    \bm{B_{\rm pol}} = B\left(\cos{\theta}\ \cos{\phi}\ \bm{\hat{i}} + \cos{\theta}\ \sin{\phi}\ \bm{\hat{j}}+\sin{\theta}\ \bm{\hat{k}}\right)\mbox{\ .}
\end{equation}

Finally, in the random field case, the three Cartesian coordinates of $\bm{B_{\rm rand}}$ are taken to be $B_\text{sat}$, given in equation~(\ref{Eq.Bsat}), where $\eta$ is now generated randomly between 0 and 1, separately in every cell. This is in contrast to the other geometries, where $\eta$ takes specific values, as discussed above (Sec.~\ref{sect2.4.2}). This mimics in a rough way the small-scale turbulence during the amplification of the field.

\begin{table}
 \caption{Notational convention for the magnetic field cases considered. The first column shows the notations used to represent a given initial field geometry (second column), with initial strength given by Equ.~\ref{Eq.Bsat} for coupling efficiency $\eta$ (third column).}
 \label{tab:abv}
 \begin{tabular}{lcc}
  \hline 
  Case & Initial field geometry & $\eta$ \\
  \hline
  PI & Poloidal & 0.5 \\
  RI & Radial & 0.5 \\
  TI & Toroidal & 0.5 \\
  UI & Uniform & 0.5 \\
  Rand & Uniform & random between 0 and 1 \\
  \hline
 \end{tabular}
\end{table}

\section{Results and discussion}
\label{sect3}

In the following, we present our results by contrasting the cases with various magnetic field configurations with a hydrodynamics-only comparison simulation, referred to as HD case. We will focus in particular on the MHD simulations with intermediate-strength fields, labeled as ``I'', for the following cases: UI for a uniform magnetic field, and RI, TI, PI denoting radial, toroidal, and poloidal field geometries, respectively. These notational conventions are summarized in Table~\ref{tab:abv}.

\subsection{Global properties of collapse}
\label{sect3.1}

By $z\sim 25$, our cosmological box contains multiple minihaloes, able to host dense baryonic cores. The lowest-mass halo has a mass of $2.3\times 10^{5}\,M_{\odot}$, and is just marginally able to trigger the collapse of primordial gas. However, we here focus on the most massive minihalo in the simulation box, of mass $M_{\rm{halo}} = 1.2\times 10^{6}\,M_{\odot} $ and virial radius $178~\rm{pc}$, as a more typical representative of a primordial star forming region.

We begin by considering the radial profile of the magnetic field magnitude for the intermediate-strength cases (top panel of Fig.~\ref{fig:comparisonB}), at the time when the number density reaches a value of $10^{12} \pcc$. As can be seen, the initial geometry has a minor effect on the final magnitude, such that the detailed field configuration has little effect on the overall star formation process. For completeness, to verify that the divergence constraint, $\bm{\nabla} \cdot \bm{B} =0$, is well satisfied in our suite of MHD simulations here, we show the radial profile of the (absolute) magnetic field divergence, expressed in dimensionless form (Fig.~\ref{fig:comparisonB}, bottom panel). It can be seen that this term is less than unity in the central region, which demonstrates that the $\bm{\nabla} \cdot \bm{B}$ error is dynamically unimportant \citep[see appendix C in][]{wang2009}. We have verified that this is the case for other evolutionary stages as well, with the exception of the random field briefly after its insertion.

The distinction between the HD and MHD cases starts to become apparent at a density of $10^{8} \pcc$, at which the magnetic field is introduced and MHD effects begin to modulate the evolution. Continuing the simulations beyond this stage to reach densities of $10^{12} \pcc$ takes a time of $15~\rm kyr$ for the HD case, and $20~\rm kyr$ for the cases with intermediate-strength magnetic fields. This demonstrates that including the magnetic fields leads to a delay of the collapse due to the presence of magnetic pressure. In Fig.~\ref{fig:timeevol}, we compare evolutionary sequences for the HD and select MHD cases, at three representative moments, quantified by the maximum densities reached at that time, namely $10^{9} \pcc$, $10^{12} \pcc$ and $10^{15} \pcc$. In each case, we find a central object surrounded by a more or less well-organized disc. The subsequent evolution engenders multiple fragmentation of the disc (middle column). For the HD case, however, the final moment reveals a radially symmetric disc with little remaining substructure around it (top right panel). In the TI case, two clumps temporarily emerge (middle panel), which merge into one central object by the end of the simulation, surrounded by a deformed disc under the influence of the toroidal magnetic field (right-most panel). Summarizing this behaviour, we find that adding the magnetic field inhibits fragmentation of the disc around the central object, with the toroidal field having the strongest effect. In other words, the field stabilizes the disc. We have verified that for the low-$\eta$ magnetic field cases, the change of geometry had little effect on the morphology of the disc. The highest-$\eta$ cases, on the other hand, for the same geometry lead to a thinner and less dense disc, which inhibits star formation. 

It is interesting to compare the HD case with the TI one for different spatial scales. Here, the toroidal initial field is particularly important, because our comparison simulation, where we self-consistently follow the magnetic field evolution from a small initial seed field (see Sec.~\ref{sect2.4.1}), results in a toroidal field component that begins to dominate at number density $10^8 \pcc$, in agreement with \citet{machida2008}. This result also agrees with the findings in \citet{sharda2021}, where they show that the $\alpha \Omega$ dynamo transforms the random small-scale field into an ordered, large-scale toroidal component within the protostellar disc. In Fig.~\ref{fig:HDvsMHDI}, we present views of the central morphology for three box sizes, 0.1\,pc, 0.01\,pc and $5\times 10^{-4}$\,pc. 
The central `protostar' defined by its photospheric surface (see Sec.~\ref{sect2.2}) is visible in the middle panel, although we point out again that this is an artificial object, resulting from our limited numerical resolution. This object appears more compact with an elongated disc in the TI case, in contrast to the more symmetric disc in the HD case. The most noticeable difference in the morphology between the HD and the MHD cases is evident in the right-most panels, exhibiting the smallest spatial scale. In the HD case, there is not a single central peak but several clumps. These features are not seen in the MHD case, where only one central peak is formed. The dynamics inside this photospheric surface, not reliably resolved here, will be investigated further in the high-resolution runs, employing a threshold number density for the artificial optical depth in equation~(\ref{Eq.tauart}), set to the much higher $n_{\rm th} = 10^{15}\pcc$ (as discussed in Sec.~\ref{sect2.2}).

In the high-resolution runs, the maximal number density reached is $n_{\rm{max}}=3\times 10^{15}\pcc$, for the HD, PI and TI cases. We thus get somewhat closer to the true protostellar stage, which would require reaching densities of $\sim 10^{21}\pcc$ \citep[see][]{greif2012}, in excess of our available computational resources. In Fig.~\ref{fig:HDvsMHDI_noopacity}, we show the density distribution for the HD run (top row), the PI run (middle row) and the TI run (bottom row), at the final time reached in each case. It is noticeable that in the HD case, a clumpy central disc forms, whereas in the PI case two fragments emerge close to the central protostar. In contrast, the TI case results in a single fragment (or protostar). This shows that the geometry of the magnetic field is important in determining its effect on the fragmentation and the morphology of the disc. This is due to the fact that the ratio of the magnetic to gas pressure, $1/\beta$, is one order of magnitude larger in the TI case than the PI case.

\subsection{Thermodynamics of collapse}
\label{sect3.2}
Further insight into the star formation process can be inferred from the corresponding thermodynamics of the collapsing gas. Particularly useful is the temperature vs. number density phase diagram, which reflects the key physical processes at work, such as the important radiative cooling and heating terms, compressional and shock heating, as well as numerical resolution effects. In Fig.~\ref{fig:comparisonTEMP}, we compare the HD with the toroidal (TI) MHD case.
The initial increase of the temperature in the range up to $n = 1\pcc$ is due to compressional heating of the collapsing gas, but a decrease follows owing to the formation of H$_2$ molecules via the H$^{-}$ channel \citep{haiman1996}. This decrease occurs after the $\rm{H}_2$ fraction reaches a value of $f_{\rm H_2}=10^{-3}$. The temperature minimum is achieved at $n = 10^4\pcc$ with $T\simeq 200-300$\,K, corresponding to what is often termed the characteristic, or `loitering' state \citep{bromm2002}. Towards higher densities, H$_2$ cooling becomes less efficient, when the rotational level populations transition to LTE values, resulting in a more gradual temperature increase due to the continuing gravitational collapse. At $n = 10^8\pcc$, the three-body reactions become effective leading to enhanced formation of H$_2$ molecules \citep{palla1983}, converting the primordial gas into a fully molecular phase. Due to the boost in cooling, another (local) temperature minimum is achieved, but the compressional heating succeeds to rise the temperature again \citep[e.g.][]{yoshida2006}. As can be seen in Fig.~\ref{fig:comparisonTEMP}, both HD and MHD (TI) cases exhibit very similar behaviour for $n < 10^8\pcc$. Beyond this number density, temperatures reach somewhat lower values in the MHD case. The reason is that magnetic pressure is countering the collapse, thus reducing the compressional heating. We note that the different geometries of the magnetic field do not noticeably affect the overall thermodynamic evolution.

The relative temperature suppression in the runs with a magnetic field can also be seen in Fig.~\ref{fig:comparisonT}, where we show the radial temperature profile for select cases. The temperature drop close to the centre, seen in all cases, is a numerical artefact due to the adopted stiffened equation of state that artificially slows down compressional heating in the central region. The temperature increases to a maximum of $\simeq 6000$\,K in the HD case, whereas in the MHD case the maximum temperature reached is $\simeq 5500 $\,K, since magnetic pressure counteracts the collapse for the MHD runs. We again note that the detailed magnetic field geometry does not greatly affect the resulting thermal profiles. The evolution beyond the numerical threshold density, $n_\text{th}$, cannot be resolved here, rendering the approach to the final hydrostatic core inaccessible \citep[see][]{yoshida2008, greif2012}.

In terms of the resulting kinematics, our simulations indicate that the infall velocity is greater in the HD case than in the MHD cases, reflecting the role of the magnetic field in slowing the collapse, as shown in Fig.~\ref{fig:comparison2}. The square of the vorticity, defined by $\omega = \bm{\nabla} \times \bm{v}$, indicates that the centre of the disc exhibits high turbulent energy, with the caveat that the central region is not reliably treated here, due to our artificial optical depth methodology. Furthermore, we note that the radial Mach number is subsonic in the HD case, whereas in the MHD cases the discs are approaching transonic conditions. Finally, we point out that the angular momentum transport is more efficient in the TI case than in the HD case, as indicated in the steep decline of the angular momentum profile (see bottom-left panel in Fig.~\ref{fig:comparison2}).

\subsection{Fragmentation properties}
\label{sect3.3}

To investigate the stability of the protostellar disc, we consider the Toomre-Q parameter \citep{toomre1964}:
\begin{equation}
\label{eqtoomreQ}
    Q = \frac{c_{\rm s}\kappa}{\pi G \Sigma} \mbox{\ ,}
\end{equation}
where $c_{\rm s}$ is the sound speed, $\kappa$ the epicyclic frequency of the disc, and $\Sigma$ the surface mass density. This parameter determines whether perturbations in a differentially rotating, gaseous disc can grow. Assuming a Keplerian disc, we can simplify $\kappa\sim \Omega$, where $\Omega\sim v_\text{rot}/r$ is the orbital frequency \citep[e.g.][]{greif2012}. Following \cite{hirano2017}, we assume the thin disc approximation, allowing us to substitute $H\simeq c_{\rm s}/\Omega$ for the scale height. Further approximating $\Sigma\sim \rho H$, we arrive at a convenient local version for the Toomre Q-parameter, given in equation (23) of \cite{hirano2017}, as follows: 
\begin{equation}
    Q_{\rm local} \sim \frac{\Omega^2}{\pi G \rho} \mbox{\ .}
\end{equation}
The regions where $Q_{\rm local}<1$ are unstable and prone to fragmentation, whereas those with $Q_{\rm local}>1$ are stable. 

The inspection of Fig.~\ref{fig:toomreslices} reveals the following. The central configurations are morphologically different. The disc in the HD case is radially symmetric, while the symmetry is broken in the MHD (TI) case, with a dominant elongated structure. In the HD case, several clumps are embedded in the disc, emerging $\sim 100$\,AU away from the central density peak (left panel). In the MHD (TI) case, on the other hand, there are two dominant clumps along the elongated structure. We conclude that the magnetic field suppresses fragmentation into low-mass clumps by increasing the pressure-support against gravity. Even for the highest initial magnetic field amplitude, such a binary clump forms instead of a massive single system. The complex asymptotic state of this ongoing sub-fragmentation, merging, and possibly ejection of fragments is beyond the reach of our current simulation, as it would require higher resolution extended for significantly longer periods of time.

\begin{table}
 \caption{The masses (M$_{\rm 1}$) of the main `protostar', defined by the photospheric surface, when the maximum density is $10^{12} \pcc$, the masses (M$_2$) at density $10^{15} \pcc$, the corresponding accretion time, $t_{\rm acc}$, and the accretion rate, $\dot{M_*}$. Displayed are the HD, toroidal, poloidal and uniform magnetic field cases, the latter for intermediate-strength normalization. The somewhat reduced accretion rates are caused by the magnetic field for all geometries considered.}
 \label{tab:masses}
 \begin{tabular}{lcccc}
  \hline
  Case & M$_{\rm 1}~(M_{\odot})$ & M$_{\rm 2}~(M_{\odot})$ & $t_{\rm acc}~(yr)$ & $\dot{M_*}~(M_{\odot}/yr)$ \\[0.5ex] 
  \hline
  HD & 17.9 & 30.6 & 72 & 0.177\\
  TI & 7.9 & 12.0 & 66 & 0.062 \\
  PI & 18.1 & 19.0 & 52 & 0.017 \\
  UI & 1.2 & 17.5 & 1020 & 0.016 \\
  \hline
 \end{tabular}
\end{table}

\begin{table}
 \caption{Mass of the main `protostar' ($M_{\rm high}$), defined by the photospheric surface, when the maximum density is $10^{15} \pcc$. Displayed are the HD, poloidal and toroidal magnetic field cases, the latter for intermediate-strength normalization. The magnetic field leads to the formation of more massive protostars.}
 \label{tab:massesH}
 \begin{tabular}{lcccc}
  \hline
  Case & M$_{\rm high}~(M_{\odot})$ \\
  \hline
  HD & 0.02 \\
  PI & 0.05 \\
  TI & 0.14 \\
  \hline
 \end{tabular}
\end{table}

\subsection{Resulting fragment masses}
\label{sect3.4}

A basic question concerning the Pop~III stars is their mass distribution, and more specifically their IMF. In Fig.~\ref{fig:comparisonMASS}, the baryonic mass enclosed in a sphere of radius 0.01\,pc is shown as a function of radius, when the maximum density is $10^{12} \pcc$, but on this scale, the enclosed mass is virtually the same when considering the higher maximum density of $10^{15} \pcc$. The mass profiles for the HD and MHD (RI) cases are similar, because the collapse in both cases proceeds in a radially symmetric fashion. The other geometries result in an elongated morphology with a smaller enclosed mass as a function of radius.

To identify the gravitationally bound cells (termed `clumps' here), we employ the core-finder routine described in \cite{smith2009}. These clumps will likely later collapse to form one or several protostars. Fig.~\ref{fig:clumps} shows snapshots of these clumps when the number density is $10^{15} \pcc$. The left panel indicates that $\sim$ 50 clumps are formed in the HD case, but only four clumps emerge in the MHD (TI) case (see right panel). We again see the effect of the magnetic field inhibiting fragmentation.

Investigating the resulting mass distribution, we consider the histogram in Fig.~\ref{fig:hist}, taken at a density $10^{15} \pcc$. In the HD case, $\sim$50 clumps emerge, with masses close to the local Jeans mass, $M_\text{J}\sim 1 M_{\odot}$, evaluated at our resolution limit of $n_\text{th}=10^{12} \pcc$. The mass distribution in this case reflects effective fragmentation as the number density in the disc continues to increase. In contrast, for the MHD cases, the mass of the central clump depends on the geometry of the magnetic field. The resulting masses are $50\,M_{\odot}$, $10\,M_{\odot}$, and $15\,M_{\odot}$ in the PI, TI and UI cases, respectively. Furthermore, the small-mass fragments close to the resolution limit are absent in all MHD cases, as the magnetic field inhibits small-scale fragmentation. We also find that stronger fields lead to higher masses.

Concerning the estimation of the mass of the main `protostar' for a given case, as far as we can extend the present calculations, the accretion rates encountered here, $\dot{M}_\ast\simeq 0.001-0.1\,M_{\odot}{\rm\ yr}^{-1}$, are compatible with the canonical values expected for primordial protostellar discs \citep[e.g.][]{greif2012}. The latter are approximately given by $\dot{M}_\ast\sim M_\text{J}/t_\text{ff}\sim c_\text{s}^3/G$, where the sound speed reflects the range of temperatures in the central disc. The protostellar extent is here defined by the photospheric surface at an optical depth of $\tau = 0.63$, as discussed in Section~\ref{sect2.2}. The results are summarized in Table~\ref{tab:masses}, where the mass growth rate depends on whether a magnetic field is present or not. We note that the accretion rates for the central clump are higher than for the secondary clumps, driving the large separation in mass between the dominant and secondary clumps. 

In contrast, the masses of the protostars in our high-resolution runs, where we transition to a stiff equation of state at a larger threshold density of $n_\text{th}=10^{15} \pcc$, are summarized in Table~\ref{tab:massesH}. Values in the MHD cases are larger than in the HD run, with the mass in the toroidal case being the largest. This effect was missed in the lower resolution runs, thus emphasizing the need for future simulations with even higher resolution.

It is useful to compare our results with those in \citet[][hereafter, P22]{prole2022}, since both studies do not self-consistently follow the amplification of the seed field and introduce an established magnetic field later in the simulation. Next to the different numerical methodologies employed (AMR vs. moving-mesh), the two studies introduce the field at different density and with different strength. In P22, a fully saturated field, corresponding to a coupling efficiency $\eta =1$, is introduced at density $10^{10}\pcc$, two orders of magnitude larger than the density used in this study, where we consider somewhat smaller efficiencies ($\eta <1$). Although P22 have thus employed the strongest magnetic field that can be achieved, they might have missed the effect of the field at earlier evolutionary stages, when its strength would have been built up as a result of the dynamo action. Among those effects may in particular be the field's impact on the morphology of the disc, as discussed in Sec.~\ref{sect3.1}\. The irregular structure seen in our HD case, as identified by the clump finder, is likely transient. With the more robust definition of a bound, permanent fragment given by the mass enclosed in the (artificial) photosphere at $n_{\rm th}$, which is more directly comparable to P22's fragment/mass estimates via sink particles, we see a different picture. Specifically, the magnetic field in our study didn't modify the number of protostars, when using the photospheric definition, as shown in Tables~\ref{tab:masses} and \ref{tab:massesH}. Similarly, in P22 the number of sink particles was not modified by the magnetic field. Moreover, the sink masses in P22 were also unchanged by the field. This is in contrast to our study, suggesting that the magnetic field leads to more massive protostars (see Table.~\ref{tab:massesH}). Finally, P22 conclude that the magnetic field is unimportant up to density $10^{8}\pcc$, while, based on our results, magnetic fields are important ingredients in shaping the environment of the Pop~III protostellar discs, and therefore the overall process of assembling the primordial IMF.

\section{Summary and Conclusions}
\label{sect4}
In the present work, we have studied the effect of a magnetic field on the structure and evolution of the discs around Pop~III stars inside a cosmological minihalo of $M_{\rm{halo}} = 1.2\times 10^{6}\,M_{\odot}$, a virial radius of $130~\rm{pc}$, and a collapse (virialization) redshift of $z\sim 25$. In a series of idealized numerical experiments, we apply different field strengths (see Sec.~\ref{sect2.4.2}) and geometries (see Sec.~\ref{sect2.4.3}), comparing the results to those obtained in a purely hydrodynamic treatment. The magnetic field was introduced in cells at a density of $10^{8} \pcc$, and the simulations were performed until a density of $10^{12} \pcc$ in all cases, as well as to a density of $10^{15} \pcc$ in the HD case and those MHD cases with intermediate strength normalization. Our main results can be summarized as follows:

\begin{enumerate}
    \item To properly assess the impact of magnetic fields on the mass of the resulting cores requires sufficient numerical resolution within our modified optical depth approach. 
    Specifically, the runs with $n_{\rm th}=10^{12}\pcc$ missed important effects, whereas in the high-resolution runs with $n_{\rm th} =10^{15} \pcc$, it is seen that the masses of the cores in the PI and TI cases are larger than in the HD case.
    \item The detailed geometry of the magnetic field configuration has a significant effect on the resulting collapse and fragmentation, as is evident in the high-resolution runs.
    \item A robust result is that a sufficiently strong toroidal magnetic field configuration inhibits fragmentation. This can be seen by comparing the clumpy, distorted disc emerging in the HD case with the solitary, more massive fragment formed in the TI case. The PI case is somewhat intermediate, with a small multiple of fragments present.
    \item The increase in the strength of the field for the same geometry delays the collapse by about $5~\rm kyr$.
    \item Our suite of numerical experiments tentatively indicates that the presence of a sufficiently strong magnetic field renders the Pop~III IMF more top-heavy.
    \item The magnetic field decreases the accretion rate onto the protostar, with possible consequences for the protostellar evolution towards the main sequence.
\end{enumerate}

As a final remark, the present investigation was done without considering the self-consistent creation and amplification of the primordial magnetic seed field (e.g. \citealt{schober2012}; \citealt{turk2012}; \citealt{sharda2020a}; \citealt{prole2022}), with the one exception of our self-consistent comparison simulation. We here follow \cite{turk2012}, who argued that even with the highest resolution that is currently achievable, it is not possible to resolve the amplification of the magnetic field caused by the turbulent small-scale dynamo. Therefore, we assumed the magnitude of the field to be a fraction of its equipartition value, which has been shown to be the expected outcome of the self-consistent dynamo process \citep[e.g.][and references therein]{Hirano2018}. The advantage of our idealized approach is that we can carry our controlled experiments, in terms of considering a wide range of field strengths and configurations.

Another reflection concerning our approach addresses the possibility of using ideal MHD during the early phase of primordial collapse, which may be justified by the high initial degree of ionization that leads to ineffective dissipation \cite{Nakauchi2019}. However, ambipolar diffusion may become important in the primordial gas when the magnetic field gets amplified to a critical strength (e.g. \citealt{schleicher2009}, \citealt{Nakauchi2019}). Hence, since the current study uses ideal MHD, the inclusion of resistive effects such as Ohmic and ambipolar diffusion is necessary in future work, especially when the collapse will be followed to higher densities. Doing so will allow more realistic conclusions on the final mass distribution of the first stars and their resulting multiplicity.

\section*{Acknowledgements}
The authors would like to acknowledge the National Council for Scientific Research of Lebanon (CNRS-L) for granting a doctoral fellowship to Cynthia R. Saad. Computations described in this work were performed using the publicly-available \texttt{Enzo} code (\url{http://enzo-project.org}), which is the product of a collaborative effort of many independent scientists from numerous institutions around the world. Their commitment to open science has helped make this work possible. All analysis was conducted using \texttt{yt} (\url{http://yt-project.org/}). The simulations were run at the AUB HPC cluster \texttt{Octopus}. C.R.S. thanks the \texttt{Enzo} community, especially John H. Wise for his highly valuable help in implementation and debugging.

\section*{Data Availability}
The simulation data underlying this article are available on request.
Please email the lead author for obtaining the data.




\bibliographystyle{mnras}
\bibliography{references} 







\bsp	
\label{lastpage}
\end{document}